\documentclass[10pt,a4,twocolumn]{article}
\RequirePackage{amsthm,amsmath,amsfonts,amssymb}
\RequirePackage[numbers,sort&compress]{natbib}
\RequirePackage[colorlinks,citecolor=blue,urlcolor=blue]{hyperref}
\RequirePackage{graphicx}

\newcommand{\stas}{\iffalse} %% set to \iffalse for plain \iftrue for STS

\stas
\startlocaldefs
\theoremstyle{plain}

\newtheorem{theorem}{Theorem}[section]

\theoremstyle{definition}

\endlocaldefs
\fi

\usepackage{tensor}
\usepackage{bm}
\usepackage{mathtools}
\usepackage{algpseudocode}
\newcommand{\beq}{\begin{equation}}
\newcommand{\eeq}{\end{equation}}
\newcommand{\dif}[2]{\frac{{\rm d} #1}{{\rm d} #2}}

\newcommand{\ildif}[2]{{\rm d} #1/{{\rm d} #2 }}

\newcommand{\ilpdif}[2]{\partial #1/{\partial #2 }}
\newcommand{\pdif}[2]{\frac{\partial #1}{\partial #2}}
\newcommand{\pddif}[3]{\frac{\partial^2 #1}{\partial #2 \partial #3}}

\newcommand{\ilpddif}[3]{\partial^2 #1/{\partial #2 \partial #3}}
\newcommand{\defn}{\begin{quote}{\bf Definition. }}
\newcommand{\edefn}{\end{quote}}
\newcommand{\thm}{\begin{theorem}}
\newcommand{\ethm}{\end{theorem}}

\newcommand{\bmat}[1]{\left ( \begin{array}{#1}}
\newcommand{\emat}{\end{array}\right )}
\newcommand{\E}{\mathbb{E}}

\newcommand{\ts}{^{\sf T}} %transposition
\newcommand{\its}{^{\sf -T}}

\newcommand{\bp}{{\bm \beta}}

\theoremstyle{definition}

\theoremstyle{plain}

\newcommand{\eps}[3]
{{\begin{center}
 \rotatebox{#1}{\scalebox{#2}{\includegraphics{#3}}}
 \end{center}}
}

\begin{document}
\stas
\begin{frontmatter}
\runtitle{Neighbourhood Cross Validation}
\title{On Neighbourhood Cross Validation}
\else
\setlength{\textwidth}{17.7cm}
\setlength{\topmargin}{-2cm}\setlength{\textheight}{22.6cm}
\setlength{\oddsidemargin}{-5mm}
\setlength{\evensidemargin}{-5mm}
\title{On Neighbourhood Cross Validation}
\fi

\stas
\begin{aug}
\author[A]{\fnms{Simon N}~\snm{Wood}\thanksref{t1}\ead[label=e1]{simon.wood@ed.ac.uk}},

%\thankstext{t1}{I thank Heike Puhlmann and Simon Trust at the Forest Research Institute Baden-W\"urttemberg, Germany for the Terrestrial Crown Condition Inventory (TCCI) forest health monitoring survey data, and Nicole Augustin for the corresponding model structure. Thanks also to  the reviewers for comments on improving the clarity of the paper.}

\address[A]{Simon Wood is Professor of Statistical Computing, School of Mathematics,
University of Edinburgh, Edinburgh, UK\printead[presep={\ }]{e1}.}

\end{aug}
\else
\author{Simon N. Wood\\School of Mathematics, University of Edinburgh, U.K.\\{\tt simon.wood@ed.ac.uk}}

\maketitle
\fi

%\title{On modelling with reduced rank smoothing splines}
%\title{Modelling the UK black smoke network daily data: generalized additive models for gigadata}
%\title{On Neighbourhood Cross Validation}
%\author{Simon N. Wood\\School of Mathematics, University of Edinburgh, U.K.\\{\tt simon.wood@ed.ac.uk}}

%\maketitle

\begin{abstract}

Many varieties of cross validation would be statistically appealing for the estimation of smoothing and other penalized regression hyperparameters, were it not for the high cost of evaluating such criteria. Here it is shown how to efficiently and accurately compute and optimize a broad variety of cross validation criteria for a wide range of regression models estimated by minimizing a quadratically penalized loss. The leading order computational cost of hyperparameter estimation is made comparable to the cost of a single model fit given hyperparameters. In many cases this represents an $O(n)$ computational saving when modelling $n$ data. The methods make it feasible, for the first time, to use  leave-out-neighbourhood cross validation to deal with the widespread problem of un-modelled short range autocorrelation, which otherwise leads to underestimation of smoothing parameters. It is also shown how to accurately quantify uncertainty in this case, despite the un-modelled autocorrelation. Practical examples are provided, including smooth quantile regression, generalized additive models for location scale and shape, and focussing particularly on dealing with un-modelled autocorrelation.

%Example models include generalized additive models for location scale and shape and smooth additive quantile regression. Example losses include negative log likelihoods and smooth quantile losses. Example cross validation criteria include leave-out-neighbourhood cross validation for dealing with un-modelled short range autocorrelation as well as the more familiar leave-one-out cross validation. For a $p$ coefficient model of $n$ data, estimable at $O(np^2)$ computational cost, the general $O(n^2p^2)$ cost of ordinary cross validation is reduced to $O(np^2)$, computing the cross validation criterion to $O(p^3n^{-2})$ accuracy. This is achieved by directly approximating the model coefficient estimates under data subset omission, via efficiently computed single step Newton updates of the full data coefficient estimates. Optimization of the resulting cross validation criterion, with respect to multiple smoothing/precision parameters, can be achieved efficiently using quasi-Newton optimization, adapted to deal with the indefiniteness that occurs when the optimal value for a smoothing parameter tends to infinity. The link between cross validation and the jackknife can be exploited to achieve reasonably well calibrated uncertainty quantification for the model coefficients in non standard settings such as leaving-out-neighbourhoods under residual autocorrelation or quantile regression. Several practical examples are provided, focussing particularly on dealing with un-modelled autocorrelation. 
\end{abstract}
\stas
\end{frontmatter}
\fi

\section{Introduction}

This paper uses {\em Neighbourhood Cross Validation} (NCV) to refer to a general cross validation criterion encompassing a wide range of existing, statistically appealing, cross validation criteria. Many such criteria were previously computable only at a cost that rendered them impractical for the selection of more than one or two hyperparameters, often requiring models to be fully refitted to multiple sub-samples of the data. Building on previous work approximating leave-one-out cross validation \citep[e.g.][]{golub.heath.wahba, craven.wahba, gu2001gacv, wood2008gacv, beirami2017approxcv, wilson2020approxcv, stephenson2020loocv, rad2020approxcv} this paper's first contribution is to show how to avoid this computational cost when applying NCV  to a broad class of quadratically penalized regression models, demonstrating how a highly accurate approximation to the criterion can be optimized with respect to multiple smoothing parameters, at the same leading order cost as fitting the model given smoothing parameters. The paper's second contribution is to show how a version of NCV \citep{chu1991MCV,roberts2017CV} can practically be used to deal with un-modelled short range residual autocorrelation, avoiding the need to know the `correct' autocorrelation model, while still performing well calibrated uncertainty quantification. Such un-modelled autocorrelation is arguably the most substantial problem frequently `swept under the carpet' in applications of smooth regression modelling.

Cross validation has been widely used for model selection, evaluation and estimation of tuning parameters for a long time \citep[e.g][]{stone74,stone77}, with a corresponding diversity of variants developed \citep[see][]{arlot2010CV}. The basic idea of repeatedly omitting  a subset of the data during model fitting, and then assessing the quality of model predictions for the subset, gives rise to: the leave-one-out cross validation underpinning several smoothing parameter estimation criteria \citep[e.g][]{craven.wahba,golub.heath.wahba,gu2001gacv,wood2008gacv}; leave-out-several cross validation used to deal with autocorrelated data \citep[e.g.][\S 6.2.2]{chu1991MCV,roberts2017CV,wood2017igam}, see also \cite{opsomer2001}; and the $k$-fold cross validation \citep[e.g.][\S 7.10]{ESL2009} often used for validation and tuning parameter selection. In principle cross validation has the advantage of considerable generality. It can be applied to almost any loss function used for estimation, unlike the marginal likelihood based methods for smoothing parameter or variance parameter estimation \citep{wahba85,wood2011}, for example. But a major problem with cross validation is computational cost: in general the model has to be refitted for each omitted subset (fold). For some special cases, such as univariate least squares spline smoothing, the cost can be reduced essentially to that of a single model fit \citep[e.g][]{elden84,hutchinson1985,deHoog1987}. But beyond the univariate least squares setting cost considerations lead to the use of approximations such as Generalized Cross Validation \citep[GCV,][]{craven.wahba,golub.heath.wahba}. Splitting the data into only a small number of folds, as in $k$-fold cross validation, is another approach that reduces cost, but optimizing such scores with respect to hyperparameters is typically quite expensive, while the choice of folds introduces a degree of arbitrariness in the results. This paper shows how cross validation criteria can be computed efficiently and to high accuracy for a wide class of penalized regression models, optimized equally efficiently with respect to hyperparameters, and exploited for uncertainty quantification, substantially extending earlier work such as \cite{beirami2017approxcv, wilson2020approxcv, stephenson2020loocv}. The methods apply equally to likelihood and non-likelihood based regular loss functions. Particular attention is paid to how this facilitates model estimation in the presence of unmodelled short-range residual autocorrelation -- an important applied issue.   

Consider $n$ data $y_i$. Suppose that we have a regression model for $y_i$, with parameters ${\bm \theta}_i$, which are in turn determined by $p$ coefficients, $\bp$, and that the fit of the model is measured by a regular loss function
$$
\sum_{i=1}^n {\cal D}(y_i,{\bm \theta}_i).
$$ 
For example $\cal D$ might be a negative log likelihood, a negative log pseudo- or quasi-likelihood, or a different loss such as the ELF loss \citep{fasiolo2021qgam} used in quantile regression. ${\cal D}$ may itself also depend on a small number of hyperparameters, but there will typically be a larger number of hyperparameters  associated with penalties on $\bp$ applied during estimation. The penalties control the smoothness of spline terms in the model, or the dispersion of random effects. 

For example, in the special case of a generalized additive model \citep{h&t90,wood2017igam} for some exponential family distributed $y_i$, ${\cal D}$ might be the corresponding deviance and ${\bm \theta}_i$ would simply be $\mu_i = \E(y_i)$, parameterized in terms of the coefficients, $\bp$, of basis expansions for the model component smooth functions of covariates. These coefficients would then be estimated to minimize
\beq
\sum_{i=1}^n {\cal D}(y_i,\mu_i) + \sum_j \lambda_j \bp\ts {\bf S}_j \bp \label{pen.loss}
\eeq    
where the penalties, $\bp \ts {\bf S}_j \bp$, measure function wiggliness. The hyperparameters, $\lambda_j$, therefore control how smooth the estimated model functions will be. In the case in which the loss is a negative log likelihood and each parameter, $\theta_i$, is modelled by its own smooth additive linear predictor, the class of models is the distributional regression models (generalized additive models for location scale and shape) of \cite{yee1996,gamlss, mayr2012gamlssboost,klein2014dr,klein2015dr,yee2015book,wood2015plig,gamlss.book} etc.

Generically the direct cross validation approach to estimating hyperparameters is as follows. For $k = 1,\ldots, m$ choose subsets $\alpha(k) \subset \{1,\ldots, n\}$ and $\delta(k) \subset \{1,\ldots, n\}$. Usually $\delta(k) \subseteq \alpha(k)$. Let ${\bm \theta}_i^{-\alpha(k)}$ denote the estimate of $\bm \theta_i$ when data with indices in $\alpha(k)$ are omitted from the estimation data. Then the {\em neighbourhood} cross validation (NCV) criterion
\beq
V = \sum_{k=1}^m\sum_{i \in \delta(k)}{\cal D}(y_i,{\bm \theta}_i^{-\alpha(k)}) \label{NCV}
\eeq
is optimized with respect to the hyperparameters. Leave-one-out cross validation corresponds to $m=n$ and $\alpha(k)=\delta(k)=k$. $K$-fold cross validation has $m=K$, $\alpha(k)=\delta(k)$ and each $\alpha(k)$ an approximately equally sized non-overlapping subset of $\{1,\ldots,n\}$, such that $\cup_{k=1}^K \alpha(k) = \{1,\ldots,n\}$. A particularly interesting case occurs when $\text{nei}(i)$ denotes some {\em neighbourhood} of $i$ determined, for example, by spatial or temporal proximity and $\alpha(k)=\text{nei}(k)$ with $m=n$ and $\delta(k)=k$. If it is reasonable to assume that there is short range residual correlation between point $k$ and points in $\alpha(k)$, but not between $k$ and $j \notin \alpha(k)$, then NCV provides a means to choose hyper-parameters without the danger of overfit that such short range (positive) autocorrelation otherwise causes \citep[e.g.][]{chu1991MCV}. Since the residual autocorrelation is not being directly modelled, then if $\cal D$ is nominally a negative log likelihood, this approach is treating it as a log (first order or marginal) pseudo-likelihood. Under short range autocorrelation the corresponding maximum pseudo-likelihood estimators for $\bm \theta$ are consistent \citep[see e.g.][]{coxreid2004,gourieroux2017pseudolikelihood}, so that the same can be arranged under the quadratic penalization of interest here. However standard asymptotic covariance matrix results for maximum pseudo-likelihood estimators would require multiple replicates of the response vector $\bf y$ and can hence not be adapted for use in this context. Uncertainty quantification has to allow for this.      

The various versions of cross validation covered by (\ref{NCV}) have a long history, but their use for routine hyper-parameter estimation has been limited by computational cost. Other than in linear special cases, the cost of exactly evaluating (\ref{NCV}) is $O(mnp^2)$ for a regression model with $O(np^2)$ estimation cost, and often $m=n$. The aim of this paper is to provide methods by which hyperparameters can efficiently and reliably  be estimated by optimization of accurate approximations to (\ref{NCV}), reducing the cost to $O(np^2)$. Uncertainty quantification is also discussed: the link between cross validation and the jackknife being useful in the cases of a non-likelihood based loss, and in the presence of residual autocorrelation.  

The paper is structured as follows. Section \ref{ncv.comp} covers efficient and accurate computation of the NCV criterion and its derivatives w.r.t. smoothing or precision parameters. Section \ref{sec.opt} covers optimization of the criterion. Enhanced computational and statistical robustness is considered in section \ref{sec.robust}, while section \ref{ARNCV} explains how NCV can overcome the deficiencies of leave-one-out cross validation in the face of autocorrelation and section \ref{sec.uq} discusses variance estimation and the link to the jackknife. Section \ref{sec.sim} provides simulation examples of performance under autocorrelation and with quantile regression, while section \ref{sec.eg} provides brief example applications to electricity load prediction, spatial modelling of extreme rainfall and spatio-temporal modelling of forest health data. 

\section{Computing the neighbourhood cross validation criterion \label{ncv.comp}}

To compute (\ref{NCV}) we need to compute the coefficient estimates on omission of the data points in $\alpha(k)$, namely $\hat \bp^{-\alpha(k)}$, which determine ${\bm \theta}_i^{-\alpha(k)}$. 
Let $\hat \bp$ denote the model coefficient values optimizing the penalized loss, and let ${\bf g}_i$ denote the gradient of the unpenalized loss for $y_i$ with respect to $\bp$ at $\hat \bp$. Let ${\bf g}_{\alpha(i)} = \sum_{j\in\alpha(i)} {\bf g}_j$. Furthermore let ${\bf H}_\lambda$ denote the Hessian of the penalized loss with respect to $\bp$ at $\hat \bp$, while ${\bf H}_{\alpha(i),\alpha(i)}$ is the Hessian of the unpenalized loss for $y_j$ such that $j \in \alpha(i)$. Note that the rank of ${\bf H}_{\alpha(i),\alpha(i)}$ is the size of the set $\alpha(i)$. The Hessian of the penalized loss based on the complement of $\alpha(i)$, at $\hat \bp$, is therefore ${\bf H}_{\lambda,\alpha(i)} = {\bf H}_\lambda - {\bf H}_{\alpha(i),\alpha(i)}$. The change in $\hat \bp$ on omission of ${\bf y}_{\alpha(i)}$ can then be approximated by taking one step of Newton's method
\beq
{\bm \Delta}^{-\alpha(i)} = {\bf H}_{\lambda,\alpha(i)}^{-1} {\bf g}_{\alpha(i)}. \label{n.step}
\eeq
That is we use the approximation $\hat\bp^{-\alpha(i)} = \hat \bp - {\bm \Delta}^{-\alpha(i)}$.
Consider the accuracy of the approximation, for a regular loss (meeting the Fisher regularity conditions, for example). To fix ideas suppose that B-spline based sieve estimators are used for the model smooth covariate effects, so that $p$ may increase with $n$, typically at a slow rate such as $p \propto n^{1/5}$ \citep[see][]{claeskens2009}. Further assume that the sample size is increasing such that the rate of increase within the support of each basis function is the same. Then ${\bf H}_{\lambda,\alpha(i)}^{-1} = O(p/n)$, while  ${\bf g}_{\alpha(i)} = \sum_{j \in \alpha(i)} {\bf g}_j = O(1)$, provided that the size of $\alpha(i)$ is not increasing with $n$. Note that as $p$ increases the number of non-zero elements of ${\bf g}_{\alpha(i)}$ is $O(1)$, so ${\bm \Delta} = O(p/n)$, and $\|{\bm \Delta}\| = O(p^{3/2}/n)$ (Euclidean norm).

Newton's method is quadratically convergent, meaning that sufficiently close to the optimum, $\bp^*$ 
$$
\|\bp^{k+1} - \bp^*\| \le M \| \bp^{k} - \bp^*\|^2
$$
for some finite positive constant $M$ \cite[see][Thm 3.5 and A.2]{nocedal.wright}. Let ${\bm \epsilon}^k = \bp^{k} - \bp^*$, and ${\bm \Delta}^k = \bp^{k+1} -\bp^k= {\bm \epsilon}^{k+1} - {\bm \epsilon}^k$. Then $\|{\bm \Delta}^k\| \le \|{\bm \epsilon}^k\| +\|{\bm \epsilon}^{k+1}\|$, and hence $\|{\bm \Delta}^k\| \le \|{\bm \epsilon}^k\| + M \|{\bm \epsilon}^{k}\|^2$. Similarly $\|{\bm \epsilon}^k\| \le \|{\bm \Delta}^k\| + M \|{\bm \epsilon}^{k}\|^2$, so that $\|{\bm \Delta}^k\| \ge \|{\bm \epsilon}^k\| - M \|{\bm \epsilon}^{k}\|^2$. So $\|{\bm \epsilon}^k\|$ and $\|{\bm \Delta}^k\|$ are of the same order. Hence if ${\bm \epsilon}^0$ is the error in $\hat \bp$ before adding $\bm \Delta$, then $ \|{\bm \epsilon}^0\| = O(p^{3/2}/n)$. The error after one Newton step is hence $\|{\bm \epsilon}^1\| = O(p^3/n^2)$, which is in turn the error in $V/n$. For cubic splines, for example, $p=O(n^{1/5})$ is often used ($p=O(n^{1/9})$ gives the optimal convergence rates for regression splines, but the higher rate arguably makes more sense under penalization). See \cite{beirami2017approxcv} and \cite{wilson2020approxcv} for similar results in the context of approximate leave-one-out cross validation and fixed $p$. 

In itself the idea of taking single Newton steps is of course not new, and is effectively what is done anytime a second  order Taylor expansion is used to derive an easier to compute approximation \cite[e.g.][]{golub.heath.wahba, craven.wahba, gu2001gacv, wood2008gacv,  rad2020approxcv}. Indeed \cite{beirami2017approxcv, wilson2020approxcv, stephenson2020loocv} have all proposed basing a leave-one-out cross validation approximation on the leave-one-out version of (\ref{n.step}). Direct computation with this has $O(np^3)$ cost in general, an improvement on the $O(n^2p^2)$ cost of refitting for each fold, but the papers do not specify how to reduce this to the $O(np^2)$ cost of a single model fit \citep[but see][ which does hint that it is possible]{beirami2017approxcv}. Without such a reduction leave-one-out cross validation is not computationally competitive with marginal likelihood or AIC type methods, and more general cross validation methods will have impractical cost as estimation criteria. Equally important, previous attempts to use (\ref{n.step}), directly, do not address the substantial practical issues of (i), what to do if ${\bf H}_{\lambda,\alpha(i)}$ is not positive definite and (ii), how to handle the case in which (\ref{n.step}) proposes parameters for which the loss is undefined or infinite. No method for practical use with general purpose modelling software, for example, can ignore these problems. The remainder of this section explains how to obtain the cost reduction, while recognizing issue (i), for a broad class of smooth statistical regression models, while allowing for efficient optimization of the resulting NCV by smooth optimization methods. Issue (ii) is addressed in section \ref{qncv}.

There are several approaches that can reduce the cost of (\ref{n.step}) to $O(p^2)$. However the choice is narrowed when we recognise that while ${\bf H}_{\lambda,\alpha(i)}$ is almost always positive definite in practice, there is no finite sample size guarantee of this. Two aspects of indefiniteness need to be considered. How to diagnose it at low cost, and what to do when it is encountered? The pragmatic approach, on encountering indefiniteness, is simply to solve (\ref{n.step}) using a method that does not require a positive definite Hessian, in order to ensure the continuity of $V$ required for stable optimization. However, if this is done, it is imperative to at least have a means to detect that the problem has occurred, so that results can be carefully checked (for example by refitting with $V$ redefined without the problematic $\alpha(k)$). 

A reasonable strategy starts from the Cholesky factorization of the full data penalized Hessian, ${\bf R}_0\ts {\bf R}_0 = {\bf H}_\lambda$. Updating of ${\bf R}_0$ to obtain the Cholesky factor of  ${\bf H}_{\lambda,\alpha(i)}$ when the latter is a low rank update of  ${\bf H}_\lambda$, has $O(p^2)$ cost, and the update process reveals whether or not the updated Hessian is positive definite \cite[][\S 6.5.4]{GvL4}. The update fails when it is not. Hence if ${\bf H}_{\lambda,\alpha(i)}$ is detected to be positive definite then its (triangular) Cholesky factor can be used to solve (\ref{n.step}) at $O(p^2)$ cost. If an indefinite ${\bf H}_{\lambda,\alpha(i)}$ is detected then an iterative method suitable for symmetric indefinite matrices could be substituted, such as a pre-conditioned MINRES algorithm \cite[][\S6.4]{paige1975symmlq,vandervorst2003}. Alternatively, in the work reported here, the Woodbury identity \citep[][\S 18.2d]{harville:1997} is used. If ${\bf H}_{\lambda,\alpha(i)}= {\bf R}_0\ts {\bf R}_0 - {\bf UU}\ts$ where $\bf U$ is $p \times k$ then 
\begin{multline}
{\bf H}_{\lambda,\alpha(i)}^{-1}=({\bf R}_0\ts{\bf R}_0 - {\bf UU}\ts)^{-1} = \\ {\bf R}_0^{-1} \{{\bf I}_p - {\bf R}_0\its {\bf U}({\bf U}\ts {\bf R}_0^{-1}{\bf R}_0\its {\bf U} - {\bf I}_k)^{-1}{\bf U}\ts {\bf R}_0^{-1} \} {\bf R_0\its} \label{woodbury}
\end{multline}
which has $O(kp^2)$ cost for $k<p$. 

\subsection{Hessian factor downdate algorithms}

In practice the down-dating of Cholesky factors is performed by repeated application of rank 1 down dates, using the Givens and hyperbolic rotation based algorithms given in \cite[][\S 6.5.4]{GvL4}. This is straightforward in simple generalized regression settings where the unpenalized Hessian is ${\bf H} = {\bf X}\ts {\bf WX}$ for some model matrix, $\bf X$, and diagonal weight matrix, $\bf W$. Each rank 1 update of ${\bf H}$ is of the form $-{\bf X}_i\ts W_{ii}{\bf X}_i$, where ${\bf X}_i$ is the $i$th row of $\bf X$ corresponding to the omission  of the $i$th observation. If any rank one update fails due to indefiniteness, the Cholesky factor before the update attempt is restored, and the failed update added to an array of failed updates. Computation with this modified Hessian can then use (\ref{woodbury}) with the partially updated Cholesky factor as ${\bf R}_0$ and the failed updates making up $- {\bf UU}\ts$. Obviously any negative $W_{ii}$ will result in an update that can not fail. 

A broad class of models to which neighbourhood cross validation can be applied are the distributional regression models in which multiple parameters of a negative log likelihood are each dependent on a separate linear predictor \citep{gamlss,gamlss.book,yee1996,yee2015book,klein2014dr,klein2015dr,wood2015plig}. Some sharing of terms between predictors is also possible. Generically there are $K$ linear predictors ${\bm \eta}^k = {\bf X}[,J_k]\bp[J_k]$ where $J_k$ is a vector of indices of the model matrix columns and coefficients involved in the $k$th predictor. The log likelihood for $y_i$ is then of the form $l(y_i,h_1(\eta^1_i), h_2(\eta^2_i),\ldots,h_K(\eta^K_i))$ where the $h_j$ are inverse link functions. Writing the second partial derivative of $l_i$ w.r.t.  $\eta_i^j$ and $\eta_i^k$ as $l^{jk}_i$ then the Hessian is computed as 
\begin{algorithmic}
\State ${\bf H} = {\bf 0}_{p \times p}$
\For{$j=1:K$} \For{$k=1:K$} 
\State ${\bf H}[J_j,J_k] += -{\bf X}[,J_j]\ts \text{diag}( l^{jk}){\bf X}[,J_k]$
\EndFor\EndFor
\end{algorithmic}
Obviously in practice the symmetry of $\bf H$ is exploited for computational efficiency. 

The down dates of the Hessian are less obvious in this case. The following algorithm breaks the computation of the dropping of the $i$th point down into rank 1 up and down dates, assuming the existence of a function {\tt cholup(R,u,up)} which updates {\bf R} to the Cholesky factor of ${\bf R}\ts {\bf R} + \{2\mathbb{I}({\tt up}={\tt TRUE})-1\} {\bf uu}\ts $, if this is possible ({\bf R} package {\tt mgcv} provides such a function, for example).  
\begin{algorithmic}
\State ${\bf R} = {\bf R}_0$; ${\bf b}={\bf 0}$
\For{$q=1:2$} \Comment 2 pass ensures up- before down-date
\For{$j=1:K$} \For{$k=(j+1):K$} 
\State $\alpha = l_i^{jk}$
\State {\bf if} $q=1$ {\bf then} $b[j] += \alpha$; $b[k] += \alpha$.
\If{$(q=1, \alpha>0)\text{ or } (q=2, \alpha<0)$}
\State ${\bf v} = {\bf 0}$
\State ${\bf v}[J_j] = \sqrt{|\alpha|}{\bf X}[i,J_j]$ 
\State ${\bf v}[J_k] += \sqrt{|\alpha|} {\bf X}[i,J_k]$
\State ${\bf R} = {\tt cholup}({\bf R},{\bf v},\alpha>0)$
\EndIf
\EndFor
\State $\alpha = l^{jj}_i - b[j]$ \Comment b[j] times block subtracted
\If{$(q=1 \text,  \alpha>0)\text{ or } (q=2, \alpha<0)$}
\State ${\bf v} = {\bf 0}$; ${\bf v}[J_j] = \sqrt{|\alpha|}{\bf X}[i,J_j]$
\State ${\bf R}_0 = {\tt cholup}({\bf R},{\bf v},\alpha>0)$
\EndIf
\EndFor
\EndFor
\end{algorithmic}  
The basic idea is that the rank one updates creating the off diagonal blocks of $\bf H$ also add unwanted contributions to the leading diagonal blocks. These extra contributions are tracked in the length $K$ vector $\bf b$, and then corrected as part of each leading diagonal block's update. The two passes of the algorithm ensure that all guaranteed positive definite updates are made before any updates that could potentially remove positive definiteness. 

As with the simpler regression models, the algorithm can be modified to deal with indefinite updates. If a rank one update fails, we simply revert to the pre-attempt $\bf R$ and store the corresponding ${\bf v} $ as a column of ${\bf U}$ to employ with (\ref{woodbury}) or an equivalent iterative solver. 

In summary, (\ref{n.step}) is computed via two triangular solves using the cheaply computed Cholesky factor of  
${\bf H}_{\lambda,\alpha(i)}$, or via (\ref{woodbury}) using the partially updated Cholesky factor and the skipped updates, $\bf U$, if indefiniteness is detected. The latter case warrants some sensitivity checking, perhaps checking the results with the offending $\alpha(k)$ and corresponding $\delta(k)$ omitted from (\ref{NCV}).      

\subsection{Derivatives of the NCV \label{sec:dncv}}

Efficient and reliable optimization of (\ref{NCV}) with respect to multiple smoothing/precision parameters is only possible if numerically exact derivatives of the NCV criterion with respect to the log smoothing parameters are available. The derivatives of $V$ w.r.t. ${\bm \rho} = \log {\bm \lambda}$ follow from $\ildif{\hat \bp}{\rho_j}$ and 
\begin{align*}
\dif{{\bm \Delta}^{-\alpha(i)}}{\rho_j} & = - {\bf H}_{\lambda,\alpha(i)}^{-1}\dif{{\bf H}_{\lambda,\alpha(i)}}{\rho_j} {\bf H}_{\lambda,\alpha(i)}^{-1} \sum_{j \in \alpha(i)} {\bf g}_j\\ 
&~~~~~~~~~~~~~~~~~~~~~~~~~~~~~~~~~~~ + {\bf H}_{\lambda,\alpha(i)}^{-1} \sum_{j \in \alpha(i)} \dif{{\bf g}_j}{\rho_j}\\
& = {\bf H}_{\lambda,\alpha(i)}^{-1} \left ( \sum_{j \in \alpha(i)} \dif{{\bf g}_j}{\rho_j} -
\dif{{\bf H}_{\lambda,\alpha(i)}}{\rho_j}
 {\bm \Delta}^{-\alpha(i)} \right )
\end{align*}
all evaluated at $\hat \bp$. Obviously $\ilpdif{{\bf H}_{\lambda,\alpha(i)}}{\rho_j} = \ilpdif{{\bf H}_{\lambda}}{\rho_j} - 
\ilpdif{{\bf H}_{\alpha(i),\alpha(i)}}{\rho_j}$ with both terms on the right hand side depending in turn on $\ildif{\hat \bp}{\rho_j}$. The latter is obtained from (\ref{pen.loss}), or its generalization with ${\cal D}(y_i,{\bm \theta}_i)$ as the loss, by implicit differentiation,
$$
\dif{\hat \bp}{\rho_j} = - \lambda_j {\bf H}_\lambda^{-1} {\bf S}_j \hat \bp.
$$
Notice how these computations add only $O(p^2)$ operations given the preceding methods for efficient computation with $ {\bf H}_{\lambda,\alpha(i)}^{-1}$. Derivatives of $V$ w.r.t. $\rho_j$ then follow by routine application of the chain rule at $O(np^2)$ computational cost in the $k$-fold or leave one out cases. Appendix \ref{sec:deriv.eg} provides expressions for $\ildif{{\bf g}}{\rho_j}$ for some specific cases.

\section{NCV optimization \label{sec.opt}}

Given the above methods for efficiently computing (\ref{NCV}) and its derivatives with respect to $\bm \rho$, the NCV criterion can be optimized with respect to $\bm \rho$. In practice this will involve nested optimization. An outer optimizer seeks the best $\bm \rho$ according to (\ref{NCV}), with each trial $\bm \rho$ in turn requiring an inner optimization to obtain the corresponding $\hat \bp$. Such nested strategies are not as computationally costly as they naively appear, because the previous iterate's $\hat \bp$ value serves as an ever better starting value for the inner optimization as the outer optimization converges. In the current context a full Newton optimization is used for the inner optimization, since just evaluating (\ref{NCV}) anyway requires computation of ${\bf H}_\lambda$. The outer optimizer choice is less obvious.  

The main practical problem is that (\ref{NCV}), and other prediction error based smoothness selection criteria, are indefinite at very high or low values of the log smoothing parameters, because the model fit is then invariant to smoothing parameter changes. In practice problems caused by very low smoothing parameter values are extremely rare, but very high smoothing parameters occur often, corresponding to the case in which a model in the null space of a smoothing penalty is appropriate. For example, when using a cubic spline term, very large smoothing parameters occur whenever a simple linear effect is supported by the data. Since such occurrences are routine, it is necessary to be able to deal with them efficiently. 

The major problem with criteria indefiniteness at large smoothing parameters is that it can easily cause an optimizer to continually increase a smoothing parameter to the point at which numerical stability is lost as the smoothing penalty completely dominates the data. This issue is particularly acute when using Newton type optimizers, where low curvature on near flat sections of the criteria  can lead to very large steps being taken. Simple steepest decent type optimizers, or conjugate gradient methods, avoid this large steps issue, but are extremely computationally inefficient for near indefinite problems. 

Full outer Newton optimization would require the exact Hessian of (\ref{NCV}), which could be used directly to detect indefiniteness, optimization then proceeding only in the strictly positive definite subspace of the smoothing parameters. But the full Hessian of (\ref{NCV}) is both implementationally tedious and somewhat computationally expensive to obtain, making quasi-Newton optimization, based only on first derivatives, a more appealing choice. The difficulty is that the approximate Hessian used by a quasi-Newton method is positive definite by construction, rendering it useless for {\em detecting} indefiniteness.

An alternative recognises that indefiniteness arises because $\hat \bp$ ceases to depend meaningfully on one or more $\rho_j$s. That is $\ildif{\hat \bp}{\rho_j} \simeq 0$. Testing a vector condition is inconvenient and liable to involve arbitrary scaling choices. However, recognising that ${\bf H}_\lambda$\footnote{There are also arguments for using the Hessian of the unpenalized likelihood instead.} is the large sample approximate posterior precision matrix for $\bp$, then a suitable test for indefiniteness can be based on the scalar test of 
$$
\dif{\hat \bp \ts}{\rho_j} {\bf H}_\lambda \dif{\hat \bp}{\rho_j} \simeq 0.
$$  
If this condition is met and $\ilpdif{V}{\rho_j}\simeq 0$ then the $j$th  element of the proposed quasi-Newton update step can be set to zero. It is readily shown that such a step is still a descent direction, and that as such its length can always be selected to satisfy the Wolfe conditions \citep[see][section 3.1]{nocedal.wright}, the second of which is sufficient to ensure that the BFGS quasi-Newton update maintains positive definiteness of the Hessian approximation. Hence an outer BFGS -- inner Newton, stabilized nested optimization approach is the one employed here. 

The $O(np^2)$  cost of each optimization step matches the equivalent for GCV or marginal likelihood \citep[e.g.,][]{wood2011}, making NCV practically competitive for smoothing parameter estimation. However, low level computational considerations still make NCV the somewhat more expensive option in practice, as discussed in Appendix \ref{sec:low-level}.

\section{More robust versions of NCV \label{sec.robust}}

Given the tendency of prediction error criteria to be more sensitive to underfit than overfit, robust versions of GCV have a long history in the context of smooth mean regression \citep[e.g][]{robinson1989rcv,vanderLinde2000,lukas2006robust,lukas2010rgcv,lukas2016}. The basic idea is to add, to the standard leave-one-out cross validation criterion, a penalty on the change in the estimate of  $\mu_i = \E(y_i)$ on omission of $y_i$ from the fit. This penalty is essentially a stability criterion, and the analyst has to choose how much weight to give the prediction error and stability terms in the robustified criterion. The choice is fundamental as neither the tendency for overfit, nor the effect of the stability penalty, vanish in the large sample limit. Often the comparison of robust and non robust cross validation results serves as a useful check of statistical stability. 

Beyond mean regression a more general notion of model stability is required. For example
$$
V_s =\sum_{k=1}^m\sum_{i \in \delta(k)}{\cal D}(y_i,{\bm \theta}_i^{-\alpha(k)}) - 
\sum_{k=1}^m\sum_{i \in \delta(k)}{\cal D}(y_i,\hat {\bm \theta}_i)
$$
measures the sum of the difference in loss between fits with and without each $\alpha(i)$ omitted. This is a measure of how sensitive the model fit is to omission of data, which is naturally on the same scale as the NCV score. Letting $\gamma-1$ denote the weight to give to $V_s$, the robust NCV criterion then becomes
$$
V_r =\gamma\sum_{k=1}^m\sum_{i \in \delta(k)}{\cal D}(y_i,{\bm \theta}_i^{-\alpha(k)}) - 
(\gamma-1)\sum_{k=1}^m\sum_{i \in \delta(k)}{\cal D}(y_i,\hat {\bm \theta}_i).
$$
The definition of $\gamma$, such that $\gamma=1$ yields the ordinary NCV criterion, corresponds to the usual usage for robustified GCV criteria. 
If the loss is a negative log likelihood and leave-one-out or $k-$fold cross validation is used, then to within an additive constant NCV is a direct estimate of the KL-divergence, and $V_s$ might be thought of as estimating a sort of `KL-stability'.

A simpler alternative is as follows. Suppose that, rather than simply being dropped, the data for a neighbourhood are replaced by data having the opposite influence on the gradient of the loss w.r.t. linear predictor, ${\bm \eta}$. This leads to a doubling of the change in gradient on perturbation of the neighbourhood, and using the same quadratic model as in the data omission case, a doubling of the step length. In short,
$$
{\bm \eta}^{[-\alpha(i)]} \leftarrow 2 {\bm \eta}^{[-\alpha(i)]} - \hat {\bm \eta}.
$$
More generally one might choose to move a different size of perturbation, leading to
$$
{\bm \eta}^{[-\alpha(i)]} \leftarrow \gamma {\bm \eta}^{[-\alpha(i)]} - (\gamma - 1)\hat {\bm \eta}.
$$
again for $\gamma>1$. The modified $
{\bm \eta}^{[-\alpha(i)]} \leftarrow 2 {\bm \eta}^{[-\alpha(i)]} - \hat {\bm \eta}.
$ is then used in (\ref{NCV}) to obtain an alternative $V_r$.

Computation of $V_r$ requires only routine modifications of the computations for $V$. 

\subsection{Finite sample computational robustness \label{qncv}}

Sometimes models are specified in such a way that some $\bp$ values are impossible. For example, using a regression model for Poisson data with a square root or identity link will predict negative expected values for some regression coefficient values. For model fitting this is not usually problematic, because a numerical optimizer can easily avoid regions of infinite log likelihood. However the proposed NCV requires taking Newton steps without checking that the loss function is finite at the step end. Indeed making such checks would break the differentiability of the NCV score on which efficient optimization relies. Further, while the high convergence rates on which the proposal relies suggest that this issue will be rare at large sample sizes, there are no guarantees at finite sample size, and in any case it only takes one problematic datum, at some point in the NCV optimization, to break the optimization.

A simple fix for this problem is to replace the loss function in the NCV criterion with a quadratic approximation about the full model fit. Consider $\cal D$ when several linear predictors are involved. For compactness denote derivatives by sub and superscripts so that ${\cal D} \indices*{*_{\hat \eta}^j} = \left . \ilpdif{{\cal D}}{\eta_j} \right |_{\hat \eta_j}$, and use the convention that repeated indices in a product are to be summed over. Suppressing the indices relating to particular observations, we then have
\begin{multline*}
\gamma {\cal D}(\eta_1,\eta_2,\ldots) - (\gamma-1){\cal D}(\hat\eta_1,\hat \eta_2,\ldots) \simeq {\cal D}(\hat\eta_1,\hat \eta_2,\ldots)  \\ + \gamma{\cal D} \indices*{*_{\hat \eta}^j} (\eta_j - \hat \eta_j) + \gamma{\cal D}\indices*{*_{\hat \eta}^j_{\hat \eta}^k}(\eta_j - \hat \eta_j) (\eta_k - \hat \eta_k)/2.
\end{multline*}
This quadratic approximation on the right hand side is then used in place of $\cal D$ in (\ref{NCV}). The resulting criterion, $V_q$ say, is always finite. Its derivatives with respect to smoothing/precision parameters rely on differentiating the quadratic approximation to yield 
\begin{multline*}
{\cal D}\indices*{*_{\hat \rho}^i} \simeq{\cal D} \indices*{*_{\hat \eta}^j} \dif{\hat \eta_j}{\rho_i} + 
\gamma {\cal D}\indices*{*_{\hat \eta}^j_{\hat \eta}^k}\dif{\hat \eta_k}{\rho_i}(\eta_j - \hat \eta_j) \\ + 
\frac{\gamma}{2}{\cal D}\indices*{*_{\hat \eta}^j_{\hat \eta}^k_{\hat \eta}^l} \dif{\hat \eta_l}{\rho_i}(\eta_j - \hat \eta_j) (\eta_k - \hat \eta_k) \\+ 
\gamma {\cal D}\indices*{*_{\hat \eta}^j}\left (\dif{\eta_j}{\rho_i} - \dif{\hat \eta_j}{\rho_i} \right )
+\frac{\gamma}{2}{\cal D}\indices*{*_{\hat \eta}^j_{\hat \eta}^k} \left (\dif{\eta_j}{\rho_i} - \dif{\hat \eta_j}{\rho_i}\right ) (\eta_k - \hat \eta_k)\\ + \frac{\gamma}{2}{\cal D}\indices*{*_{\hat \eta}^j_{\hat \eta}^k}(\eta_j - \hat \eta_j) \left (\dif{\eta_k}{\rho_i} - \dif{\hat \eta_k}{\rho_i}\right )
\end{multline*}
which, on collection of terms, is
\begin{multline*}
{\cal D}\indices*{*_{\hat \rho}^i} \simeq 
{\cal D} \indices*{*_{\hat \eta}^j} \left \{(1-\gamma)  \dif{\hat  \eta_j}{\rho_i} + \gamma \dif{ \eta_j}{\rho_i}  \right \} \\+ 
\frac {\gamma}{2}{\cal D}\indices*{*_{\hat \eta}^j_{\hat \eta}^k} \left \{ \left (\dif{\hat \eta_k}{\rho_i} + \dif{\eta_k}{\rho_i} \right )(\eta_j - \hat \eta_j) \right . \\\left . ~~~~~~~~~~~~~~~~~~~~ + \left (\dif{\eta_j}{\rho_i} - \dif{\hat \eta_j}{\rho_i}\right ) (\eta_k - \hat \eta_k) \right \} \\ +
\frac{\gamma}{2}{\cal D}\indices*{*_{\hat \eta}^j_{\hat \eta}^k_{\hat \eta}^l} \dif{\hat \eta_l}{\rho_i}(\eta_j - \hat \eta_j) (\eta_k - \hat \eta_k) 
\end{multline*}
$V_q$ is obviously never needed when the likelihood is finite for all finite $\bp$ values, and even when this is not guaranteed seems to be needed only rarely. However there are cases where it is essential. For example, for distributional regression with the generalized extreme value distribution it is very easy to find cases for which optimization of the original $V$ fails. Again optimization of this version of the criterion proceeds exactly as for (\ref{NCV}) itself. 

\section{Neighbourhood cross validation and non-independent data \label{ARNCV}}

As mentioned in the introduction, a particularly interesting application of NCV is to the case of short range autocorrelated data. The theoretical motivation for use of cross validation is that it estimates the prediction error, or predictive loss; that is the error or loss expected when the estimated model is used to predict a new datum, independent of the fit data. Leave-one-out cross validation (LOOCV) is particularly appealing in this context, as the omission of only one datum at a time from estimation minimizes both the variance increase and any bias increase caused by the reduction in the estimation sample size. Note that \cite{bates2024cv} have shown, in  the context of linear models, that LOOCV estimates the prediction error, not on the training data set to hand, but rather over other unseen training data sets drawn from the same population. If the loss is proportional to a negative log likelihood then LOOCV directly estimates the model dependent part of the Kullback-Leibler divergence between the estimated model and the truth, which we would usually like to be as small as possible \citep[see e.g. the demonstration of the asymptotic equivalence of LOOCV and AIC in ][]{stone77}. 

LOOCV obviously fails as a prediction error/loss estimator for autocorrelated data, since in that case each datum omitted is correlated with the estimation data, rather than being independent of them: we are not estimating the error/loss for predicting {\em independent} new data. Under positive autocorrelation this means that the prediction error will be underestimated for a model that overfits, and overly complex models will tend to be selected. As a simple example consider $y_t = \alpha + e_t$, $t=1,\ldots,n$, where $e_t = \sum_{j=t-1}^{t+1} \epsilon_j$, the $\epsilon_j$ being i.i.d. Gaussian. Suppose that the mean of $y_t$ as a function of $t$ is to be estimated by a running mean of $y_{\max(1,t-k)}, \ldots, y_{\min(n,t+k)}$. Clearly the $k$ minimizing prediction error is $\ge n-1$. In contrast the $k$ minimizing LOOCV is 1 (at least for large $n$).  

The natural solution \citep[e.g.][]{chu1991MCV} is to omit sufficient neighbouring points, from the estimation data for the $i$th cross validated prediction, that the datum to be predicted is again independent of the estimation data. That is to again estimate the loss when predicting {\em independent} data, using the NCV criterion  
$$
V = \sum_{i=1}^n {\cal D}(y_i,{\bm \theta}_i^{-\text{nei}(i)})
$$
where $\text{nei}(i)$ is a set of points neighbouring point $i$, chosen so that $y_i$ is independent of ${\bf y}_{-\text{nei}(i)}$ (conditional on the regression model covariates). 

In the case of autocorrelation induced by a finite moving average process, as in the simple example above, it is possible to completely restore independence of the predicted and estimation data in this way. In other cases, such as autoregressive autocorrelation, then complete elimination of dependence will not be possible, and a decision must be made about how much residual autocorrelation to tolerate between the estimation and predicted data. In a time series application one might choose neighbourhoods large enough to eliminate `significant' autocorrelations from a residual ACF, for example.           

\section{Cross validated uncertainty quantification and the jackknife \label{sec.uq}} 

The link between cross validation and the jackknife is obvious, but is worth exploring for the NCV criterion based on predicting each single datum in turn (i.e. $\delta(k)=k $ for $k=1,\ldots,n$). Define $n \times p $ matrix $\bf D$ with $i$th row given by ${\bm \Delta}^{-\alpha(i)}(n-|\alpha(i)|)^{1/2}/(n|\alpha(i)|)^{1/2}$, where ${\bm \Delta}^{-\alpha(i)}$ is from (\ref{n.step}). Then the Jackknife estimate of the covariance matrix of $\hat \bp$ is ${\bf V}_J ={\bf D}\ts{\bf D}$ \citep[e.g.][]{shi88}. \cite[][\S 2.8.1]{davison1997bootstrap} point out that for an estimator with convergence rate $O(n^{-c})$ the power of 1/2 in the definition of $\bf D$ should really be $c$:  between $2/5$ and $4/9$ for cubic penalized regression splines, for example. But this correction turns out to be of negligible importance given what follows.   

%% p56 Davison and Hinkley argues that sqrt is wrong - would be power 2/5-4/9 for cubic regression spline.

Unfortunately the jackknife estimator only gives well calibrated inference for independent response data. When there is residual autocorrelation then the estimate is better calibrated than standard results assuming independence, but performance is still far too poor for practical use. This behaviour is easy to understand, by expanding $\hat \bp({\bf y})$ about ${\bm \mu} = E({\bf y})$
\begin{multline}
\hat \bp({\bf y}) = \hat \bp({\bm \mu}) + \sum_i \left . \pdif{\hat \bp}{y_i} \right |_{\bm \mu} ({y_i} - {\mu_i}) \\+ 
\frac{1}{2} \sum_{ij} \left .\pddif{\hat \bp}{y_i}{y_j} \right |_{\bm \tilde {\bf y}} ({y_i} - {\mu_i})({y_j} - {\mu_j}) \label{beta.exp}
\end{multline}
where $\tilde {\bf y}$ lies on a line between $\bm \mu$ and $\bf y$. In the case of a discrete $y_i$ this expression should be understood as being based on estimation using quasi-likelihood. In that case estimates are identical whether the $y_i$ are viewed as observations of discrete or continuous random variables, and the required derivatives exist. For the moment assuming $\bp$ of fixed dimension, then typically $\ilpdif{\hat \bp}{y_i} = O(n^{-1})$ and $\ilpddif{\hat \bp}{y_i}{y_j} = O(n^{-2})$, as can readily be checked for single parameter exponential family distributions, or quasi-likelihoods, used with GLMs for example. Since $E(y_i-\mu_i)=0$, and assuming bounded variance of $y_i-\mu_i$, the summations can be viewed as random walks. For independent $y_i$ it then follows that $\sum_i \left . \ilpdif{\hat \bp}{y_i} \right |_{\bm \mu} ({y_i} - {\mu_i}) = O_p(n^{-1/2})$, and $\sum_{ij} \left .\ilpddif{\hat \bp}{y_i}{y_j} \right |_{\bm \tilde {\bf y}} ({y_i} - {\mu_i})({y_j} - {\mu_j}) = O_p(n^{-1}) $. %For correlated $y_i$ the rates may be lower, but the linear term still dominates the quadratic term asymptotically.

Concentrating on the dominant linear term, while writing $\hat \bp$ for $\hat \bp({\bf y})$ and $\bp$ for $\hat \bp({\bm \mu})$ it follows that
\begin{multline*}
\hat \bp^{-\alpha(i)} \simeq \bp + \sum_{j \notin \alpha(i)}\pdif{\bp}{ y_j} ( y_j - \mu_j) \\ \Rightarrow 
{\bm \Delta}^{-\alpha(i)} = \hat \bp^{-\alpha(i)} - \hat \bp \simeq  \sum_{j \in \alpha(i)}\pdif{\bp}{ y_j} ( y_j - \mu_j).
\end{multline*}
Hence while
\begin{align}
\begin{split}
{\bf V} &= \E \{(\hat \bp - \bp)(\hat \bp - \bp)\ts \} \\ &= \E \left \{ \pdif{\bp}{\bf y} ({\bf y} - {\bm \mu})({\bf y} - {\bm \mu})\ts \pdif{\bp \ts}{\bf y\ts}   \right \} + O(n^{-3/2}) 
\end{split}\label{cov.mat}
\end{align}
the components of the Jackknife estimator have expectation
\begin{multline*}
\E \{ (\hat \bp^{-\alpha(i)} - \hat \bp)(\hat \bp^{-\alpha(i)} - \hat \bp )\ts \} \\ \simeq 
\E \left \{ \pdif{ \bp}{{\bf y}_{\alpha(i)}} ({\bf y}_{\alpha(i)} - {\bm \mu}_{\alpha(i)})({\bf y}_{\alpha(i)} - {\bm \mu}_{\alpha(i)})\ts \pdif{\bp \ts}{{\bf y}_{\alpha(i){\sf T}}}   \right \}.
\end{multline*}
So the Jacknife estimator partly accounts for residual correlation in $\bf y$, but its component terms ignore the correlation between ${\bf y}_{\alpha(i)}$ and  ${\bf y}_{-\alpha(i)}$, and are hence missing part of the correlation structure expected even if each $y_i$ is only correlated with terms in neighbourhood $\alpha(i)$.

To avoid the problem of neglected residual autocorrelation, a simpler estimator of the coefficient covariance matrix is based on the observed version of (\ref{cov.mat}). Let $b_{ki} = \ilpdif{\beta_k}{y_i}=O(n^{-1})$ and $e_i=y_i-\mu_i$ be the $i$th residual. Reusing the assumption of no residual correlation between $y_i$ and ${\bf y}_{-\alpha(i)}$ (and assuming that $j\in \alpha(i) \Leftrightarrow i \in \alpha(j)$), then (\ref{cov.mat}) can be re-written element-wise as $$ 
nV_{km} = n \sum_i^n b_{ki}\sum_{j \in \alpha(i)} b_{mj} \E(e_ie_j)+ O(n^{-1/2})
$$   
with observed version 
$$ 
n \hat V_{km} = n \sum_i^n b_{ki}\sum_{j \in \alpha(i)} b_{mj} e_ie_j.
$$   
Then
\begin{align*}
n(\hat V_{km} -V_{km}) &= n \sum_i^n b_{ki}\sum_{j \in \alpha(i)} b_{mj} \{e_ie_j - \E(e_ie_j)\}\\ &= O_p(n^{-1/2}),
\end{align*}  
since the $\sum_j$ term is a zero mean $O_p(n^{-1})$ random variable with bounded variance, provided $|\alpha(i)|$ does not grow with $n$, and $\sum_i$ is then over a random walk. Clearly if $|\alpha(i)|$  grows with $n$ the convergence rate will be lower, with convergence failure for  $|\alpha(i)| = O(n)$. Notice that these requirements on $|\alpha(i)|$, while targetting a fixed $\bp$, amount to requiring a separation of scales between the signal and the residual autocorrelation scale length.  

Computation is especially simple if we note that from application of (\ref{beta.exp}) to the leave-one-out case we can write ${\bm \Delta}^{-i} = \left . \ilpdif{ \hat \bp}{ y_i} \right |_{\hat \mu_i} ( y_i - \hat \mu_i) + O_p(n^{-2})$, so that the estimated version of (\ref{cov.mat}) becomes:
\beq
\hat {\bf V} = 
\sum_i {\bm \Delta}^{-i}  \sum_{j \in \alpha(i)} {{\bm \Delta}^{-j}}\ts. \label{basic.cov}
\eeq
To improve finite sample performance this could arguably be scaled by ${n}/(n-\tau)$ where $\tau$ denotes the (effective) degrees of freedom of the fitted model. But unfortunately the estimator anyway substantially underestimates variances at finite sample sizes. This can be understood by considering the residual products involved in (\ref{basic.cov}). Denote the true residual product by ${\bf a} = (y_i-\mu_i)({\bf y}_{\alpha(i)}-{\bm \mu}_{\alpha(i)})\ts$, and let $\hat \mu_i = \mu_i + \hat \delta_i$, where $\hat \delta_i$ is the error in $\hat \mu_i$.
\begin{align*}
(y_i - \hat \mu_i)({\bf y}_{\alpha(i)} - \hat {\bm \mu}_{\alpha(i)})\ts & \\= 
 {\bf a}- (y_i-\mu_i)\hat{\bm \delta}_{\alpha(i)}\ts  & -\hat \delta_i ({\bf y}_{\alpha(i)}-{\bm \mu}_{\alpha(i)})\ts  + \hat \delta_i \hat {\bm \delta}_{\alpha(i)}\ts \\
 = 
 {\bf a}- (y_i-\hat \mu_i)\hat{\bm \delta}_{\alpha(i)}\ts  & -\hat \delta_i ({\bf y}_{\alpha(i)}-\hat {\bm \mu}_{\alpha(i)})\ts  - \hat \delta_i \hat {\bm \delta}_{\alpha(i)}\ts.
\end{align*}
The difficulty arises because $\hat \delta_i$ tends to be positively correlated with the residuals in its neighbourhood when these are positively correlated with each other (a cluster of positive residuals leads to a positive error, for example). Consider the contribution that the first error term above makes to the summations involved in $\hat V_{km}$, namely 
$$
-n\sum b_{ki} (y_i - \hat \mu_i) \sum_{j\in \alpha(i)}b_{mj} \hat \delta_j.
$$
$\hat \delta_j$ is at best $O_p(n^{-1/2})$. The correlation between $y_i-\hat \mu_i$ and the $\hat \delta_j$ here means that the $i$-summation is over terms that have positive expectation, so that the error term is then itself $O_p(n^{-1/2})$. Were $y_i-\hat \mu_i$ uncorrelated with $\hat \delta_j$ then the $i$-summation would again be a random walk and the error term $O_p(n^{-1})$. In fact in the penalized regression setting this effect not only adds downward bias of the order of the error, it actually worsens the convergence rate, as outlined in Appendix \ref{sec:pen-var}. 

So to reduce the impact of this bias on $\hat {\bf V}$ to negligible order we need to replaced the residuals $\hat e_i = y_i - \hat \mu_i$ with residuals constructed from a model not estimated using ${\bf y}_{\alpha(i)}$. The cross validated residuals, $\tilde e_i = y_i - \hat \mu^{-\alpha(i)}$ are the obvious candidate, and to use them in place of the $\hat e_i$ simply define 
$$
\tilde{\bm \Delta}^{-k} = {\bm \Delta}^{-k}{\tilde e_k}/{\hat e_k},
$$      
and use it in place of $ {\bm \Delta}^{-k}$ in (\ref{basic.cov}) (call the result $\tilde {\bf V}$).
%The correction terms on the right hand side can be computed if $\hat \delta_i$ can be estimated. The obvious way to do this is to compute the difference between $\hat \mu_i$ and a best estimate of $\mu_i$ of the form $\bar \mu_i = \gamma \hat \mu_i + (1-\gamma) \tilde \mu_i$ for some $\gamma \in (0,1)$, where  $\tilde \mu_i = \hat \mu^{-\alpha(i)}_i$. Then $\hat \delta_i \approx \hat \mu_i - \bar \mu_i = (1-\gamma)(\tilde \epsilon_i - \hat \epsilon_i )$ where $\tilde \epsilon_i = y_i - \tilde \mu_i$. Defining ${\bm \Xi}^{-k} = {\bm \Delta}^{-k}\hat \delta_k/\hat \epsilon_k $ the corrected version of (\ref{basic.cov}) becomes
%\beq
%\hat {\bf V} = %\frac{n}{n-\tau}
%\sum_i ({\bm \Delta}^{-i} + {{\bm \Xi}^{-i}})  \sum_{j \in \alpha(i)} ({{\bm \Delta}^{-j}}\ts+{{\bm \Xi}^{-j}}\ts).\label{ncv.cov}
%\eeq
%If $\tilde \mu_i$ and $\hat \mu_i$ are treated as independent with the same variance then $\gamma = 1-1/\sqrt{2}\approx 0.3$ eliminates the shrinkage bias in $\hat \delta_i $ resulting from $\hat \mu_i$ being used in $\bar \mu_i$, while $\gamma = 0.5$ corresponds to no correction: $\gamma = 0.4$ was used here. 

An immediate generalization replaces the simple residuals used above with generalized residuals based on the loss itself and constructed such that zeros for all residuals would imply $\hat \bp = \bp$, so that (\ref{beta.exp}) can be replaced by an expansion (about 0) directly in terms of these residuals. Deviance residuals can be used, for example.

So we have three distinct cases: 
\begin{enumerate}
\item For independent data and a general loss, the jackknife covariance matrix estimator  can be used.
\item For data subject to unmodelled short range autocorrelation (and sufficiently large $n$) then (\ref{basic.cov}) is used, but based on $\tilde{\bm \Delta}^{-k}$.
\item Using a negative log likelihood with independent data, we can use the standard large sample approximation $\bp | {\bf y} \sim N({\hat \bp, {\bf H}_\lambda}^{-1})$.
\end{enumerate} 
1 and 2 result in frequentist covariance matrices, which do not account for smoothing bias, and are therefore likely to result in miscalibrated inference. In case 3 this problem is largely overcome by the use of the Bayesian posterior covariance matrix of the model coefficients. The Bayesian covariance matrix can be interpreted as the frequentist covariance matrix plus a Bayesian bias correction: the expectation, under the prior, of the squared bias matrix \citep[see in particular][]{nychka88,marra.wood2012}. 

Computation of an appropriate Bayesian bias correction for a general regular loss can be conducted under the coherent belief updating framework of \cite{bissiri16}. Under this framework prior beliefs are updated to posterior beliefs using a general negative loss in place of the usual log likelihood, but doing so requires an estimate of the {\em learning rate} parameter, $\nu$, which is the weighting of the loss relative to the prior in the belief updating framework. In the current context it is hence a multiplier on the unpenalized Hessian of the loss. Obviously NCV and corresponding point estimates are insensitive to $\nu$ -- any multiplicative change in $\nu$ simply results in the same change in the estimated $\bm \lambda$. However, the asymptotic Bayesian and frequentist covariance matrices, ${\bf V}_b={\bf H}_\lambda^{-1}$ and ${\bf V}_f ={\bf H}_\lambda^{-1}{\bf H}{\bf H}_\lambda^{-1}$ are not invariant, instead scaling directly as $\nu^{-1}$. Hence $\nu$ can be estimated from the relative scaling of ${\bf V}_f$ and $\hat {\bf V}$ (e.g. from the ratio of their traces), and the resulting $\hat \nu$ used to compute the Bayesian bias correction, $({\bf V}_b - {\bf V}_f) \hat \nu^{-1}$, to be added to $\tilde {\bf V}$.

%Uncertainty estimates for the log smoothing parameters can also be obtained. Let $\bf B$ be the approximate inverse Hessian matrix at convergence of the BFGS optimization. Then (\ref{basic.cov}) provides a direct estimate for the the covariance matrix of $\bm \rho$ upon substitution of ${\bm \Delta}^{-i}$ by ${\bf Bg}^{-i}$ .

% and $\bf P$ be the matrix with ith row ${\bf B}{\bf g}^{-\alpha(i)}(n-|\alpha(i)|)/(n|\alpha(i)|)$. Then the jackknife estimated covariance matrix for $\bm \rho $ is ${\bf V}_\rho = {\bf P}\ts {\bf P}$.

\section{Example simulations \label{sec.sim}}

\begin{figure*}
\eps{0}{.8}{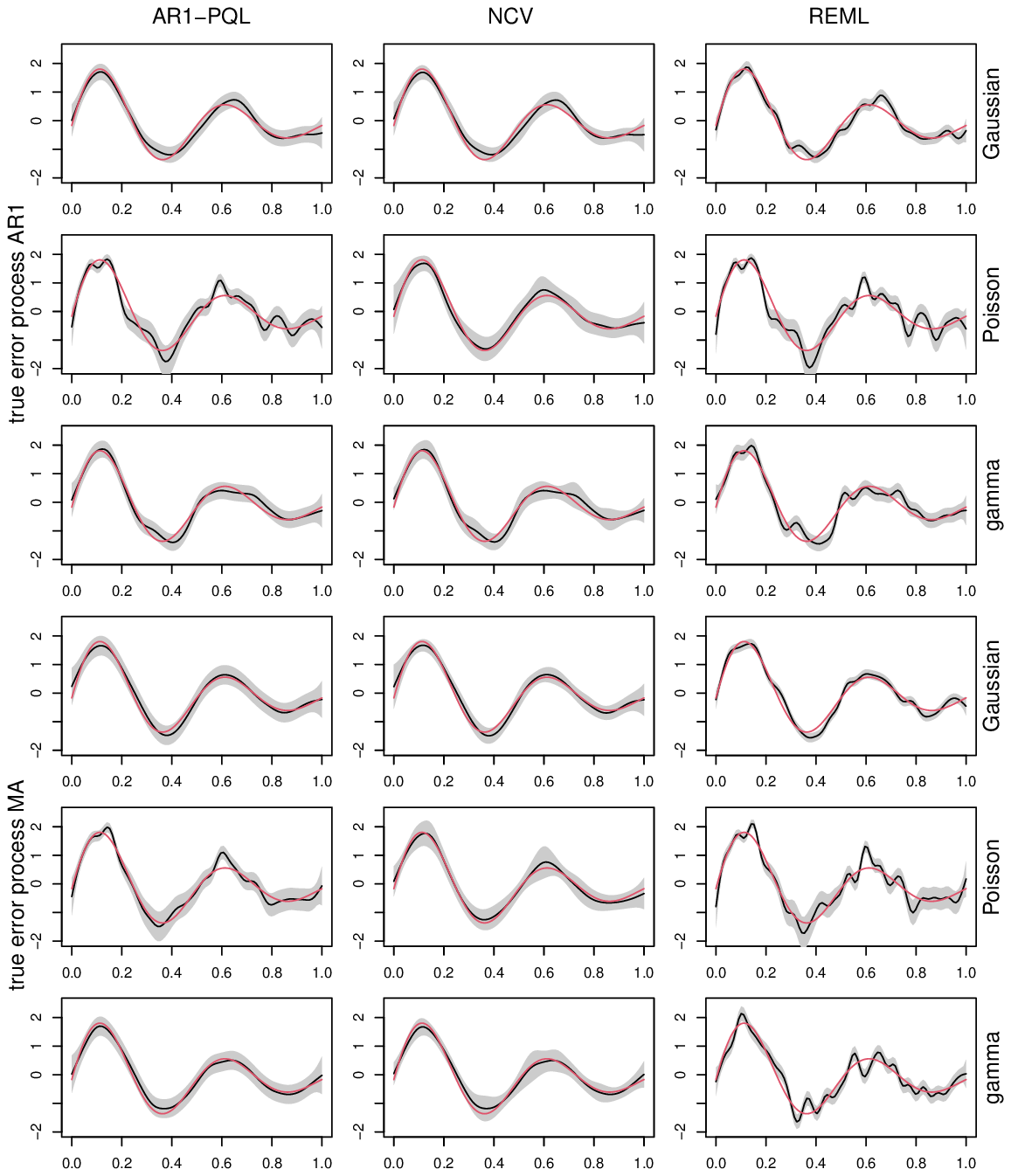}
\caption{Reasonably typical example smooth estimates using PQL with working AR1 model (left),  NCV (middle) and REML (right). In each panel nominal 95\% confidence bands are shown in grey, truth is in red. For the top three rows autocorrelation was generated using an AR1 process, and in the bottom three an MA process. The top three rows are for Gaussian, Poisson and gamma responses respectively and the same ordering is repeated for the bottom 3 rows.}\label{AR-sim}
\end{figure*}

By way of illustration of the ability of NCV to cope with short range autocorrelation, $x,y $ data were simulated from the function $f(x) = 2.5\sin(4 \pi x)\exp(-2x)$ evaluated at $x$ values spaced equally over $[0,1]$. Gaussian, Poisson and gamma simulations were run, with short range autocorrelation generated from Gaussian AR or MA processes. In the AR1 case the autocorrelation model was $e_{i+1} = 0.6 e_i + \epsilon_i$, where $\epsilon_i $ are i.i.d $N(0,1)$ deviates. In the MA case the model was $e_i = \sum_{j=i-2}^{i+2} \epsilon_j$, but linearly scaled so that $\text{var}(e_i) = 1$. Then three response models were used: $y_i = f(x_i) + e_i$, 
$y_i \sim \text{Poi}\{ \exp(f(x_i)+e_i-\sigma^2_e/2)\}$ or  $y_i \sim \text{gamma}\{\exp(f(x_i)+e_i-\sigma^2_e/2),\phi\}$ where $\phi=0.1$. 

The data were then fitted using appropriate spline models with the autocorrelation handled in one of four ways. The first option used penalized quasi-likelihood based fitting, with an AR1 correlation model on the working model scale. This is the correct model only for the Gaussian AR1 simulation, but is typical of what is often done in applications to somewhat account for otherwise un-modelled autocorrelation. The second option used NCV in which $\delta(i)=i$ and $\alpha(i)$ consisted of $i$ and its 8 nearest $x$-neighbours. The final options simply ignored the autocorrelation and estimated smoothing parameters using REML or GCV. Figure \ref{AR-sim} shows reasonably typical reconstructions (the second replicate of each case). Table \ref{AR-sim-res} gives coverage probability and MSE results. NCV generally outperforms the alternatives by some margin, with coverage probabilities close to nominal at larger sample sizes for the moving average correlation process for which it is ideally suited. It is of course not a panacea: when there is some remaining autocorrelation between each point $i$ and the points outside $\alpha(i)$, then coverage probabilities are somewhat degraded.

\begin{table*}
 \begin{center}
 \begin{tabular}{c|cccccccccccc}
 &\multicolumn{6}{c}{AR1 correlation}&\multicolumn{6}{c}{Moving average correlation}\\
 &\multicolumn{3}{c}{$n=250$}&\multicolumn{3}{c}{$n=1000$}&\multicolumn{3}{c}{$n=250$}&\multicolumn{3}{c}{$n=1000$}\\
 &Gau & Poi & $\Gamma$ &Gau & Poi & $\Gamma$&Gau & Poi & $\Gamma$ &Gau & Poi & $\Gamma$\\
 \hline
 S/N & 1.14 & 0.76 & 0.87 & 1.14 & 0.73 & 0.79 & 1.44 & 0.89 & 1.07 & 1.42 & 0.87 & 0.97 \\
 NCV CP & .923 & .902& .909 & .943 & .937 & .926 & .938 & .925 & .929 & .957 & .949 & .950\\
 PQL-AR CP & .744 & .772 & .831 & .965 & .779 &   .930 & .610 & .820 & .728 & .994 & .840 & .959 \\
 REML CP & .643 & .741 & .676 & .695 & .709 & .733 & .470 & .773 & .562 & .617 & .752 & .679\\
 GCV CP & .610 & .748 & .635& .666 & .693 & .706 & .453 & .774 & .533& .593 & .734 & .649 \\
 NCV MSE & .085 & .130 &  .103 & .024 &  .036 & .030 &  .068 & 
  .105 &  .080 & .019 &  .030 & .022\\
 PQL-AR MSE & .297 & .197 & .166 & {\bf .022} & .059 & {\bf .029} & .611 & .142 & .322 & {\bf .018} & .042 & {\bf .022}\\
 REML MSE & .159 & .237 & .174 & .046 & .077 & .058 & .194 & .191 & .190 & .046 & .060 & .051 \\
 GCV MSE & .241 & .629 & .268 & .072 & .124 & .085 & .223 & .599 & .248 & .061 & .098 & .070\\
 \hline
 \end{tabular}
\end{center}
\caption{Nominal 95\% confidence interval coverage and mean squared error for penalized regression splines with smoothness selection by GCV, REML and NCV with $\alpha(i)$ consisting of the 4 neighbours of $i$ on either side plus $i$ itself. The same spline was also fitted using PQL with an AR1 residual correlation model on the working model scale. Notice that NCV achieves close to nominal coverage for the larger sample size and MA process, for which it is eliminates all residual autocorrelation effects in smoothness selection.  Cases in which a non-NCV method outperformed NCV are highlighted in bold. For the most part NCV achieves better coverage probabilities and lower MSE than the PQL approach, even for the Gaussian - AR1 simulations for which the latter model is correct. Results are over 500 replicates.} \label{AR-sim-res}
\end{table*}

\begin{figure*}
\eps{-90}{.7}{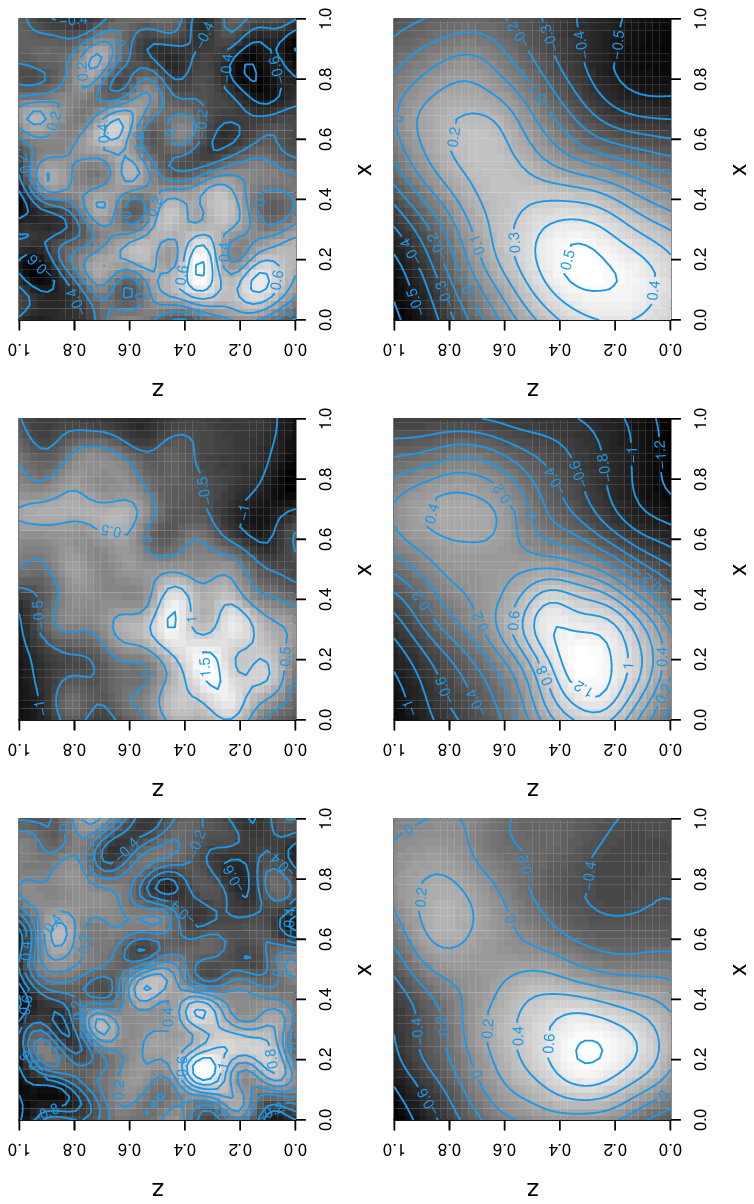}
\caption{Typical example smooth estimates using REML (top) and NCV (bottom) smoothing selection, when spatial data are autocorrelated, and the mean field is a smooth sum of two Gaussians. Data are simulated on a grid, with errors that include a moving average process, where the average is over each grid point and its 8 nearest neighbours. NCV omits each point and its (24) autocorrelated neighbours in turn. Left is for Gaussian response, middle Poisson and right gamma.}\label{spatial-sim}
\end{figure*}

\bigskip

\noindent {\bf Spatial example.} Data were also simulated from a true function, $f(x,z)$, that was the sum of two scaled Gaussian p.d.f.s over a square $x,z$ grid, with equal grid spacing in $x $ and $z$. For each grid point independent $N(0,1)$ deviates $\epsilon^\prime$ were simulated. Dependent deviates $\epsilon$ were then produced for each grid point by averaging $\epsilon^\prime$ values for the grid point and its 8 nearest neighbours. The average was weighted: 1 for the point itself, 0.5 for each of the 4 nearest neighbours and 0.3 for the 4 next-nearest neighbours. Edges were dealt with by simply extending the grid for the $\epsilon^\prime$, so that the required components of $\epsilon$ were always available. 3 models were simulated from: the Gaussian model $y_i = f(x_i,z_i) + \epsilon_i$, the Poisson model $y_i \sim \text{Poi}\{\exp(f(x_i,z_i)+\epsilon_i-\sigma^2_\epsilon/2)\}$ and the gamma model $y_i \sim \text{gamma}\{\exp(f(x_i,z_i)+\epsilon_i-\sigma^2_\epsilon/2),\phi\}$ where $\phi = 0.1$. Grids of size $18 \times 18$ and $36 \times 36$ were used. 

Models were then estimated in which the smooth functions of $x $ and $z$ were represented using penalized thin plate regression splines, with basis dimensions of 100 for the smaller grid and 150 for the larger. Smoothing parameters were estimated by GCV, REML and by NCV with the omitted neighbourhoods for each point consisting of the square of 25 points centred on the point. Typical reconstructions using REML and NCV, for the larger grid, are shown in Figure \ref{spatial-sim}. The tendency of REML to severely over fit when there is un-modelled short range residual autocorrelation is shown even more strongly by GCV \citep[see][for an explanation]{krivobokova2007}. Performance of the model fits over 500 replicates is shown in table \ref{spatial-sim-res}.

\begin{table*}
 \begin{center}
 \begin{tabular}{c|cccccc}
 &&$18 \times 18$&&&$36 \times 36$&\\
 &Gaussian & Poisson & gamma &Gaussian & Poisson & gamma\\
 \hline
 S/N & 0.95 & 1.11 & 0.68 & 0.94 & 1.08 & 0.66 \\
 NCV CP &  .939 & .962 & .934 & .950 & .952 & .948\\
 REML CP & .321 & .827 & .671 & .467 & .763 & .672 \\
 GCV CP & .319 & .826 & .653& .459 & .749 & .651 \\
 NCV MSE & .040 & .064 & .037 & .012 & .021 & .013 \\
 REML MSE & .111 & .093 & .091 & .063 & .061 & .056 \\
 GCV MSE & .114 & .116 & .110 & .068 & .082 & .069 \\
 \hline
 \end{tabular}
\end{center}
\caption{Spatial moving average simulation example results over 500 replicates, for two grid sizes and 3 response distributions, reporting signal to noise ratio (ratio of standard deviation of $f$ to standard deviation of residual), coverage probabilities of nominally 95\% CIs and mean square error, for NCV, REML and GCV smoothing parameter estimation.   \label{spatial-sim-res}}
\end{table*}

\bigskip 

\begin{figure}
\eps{-90}{.55}{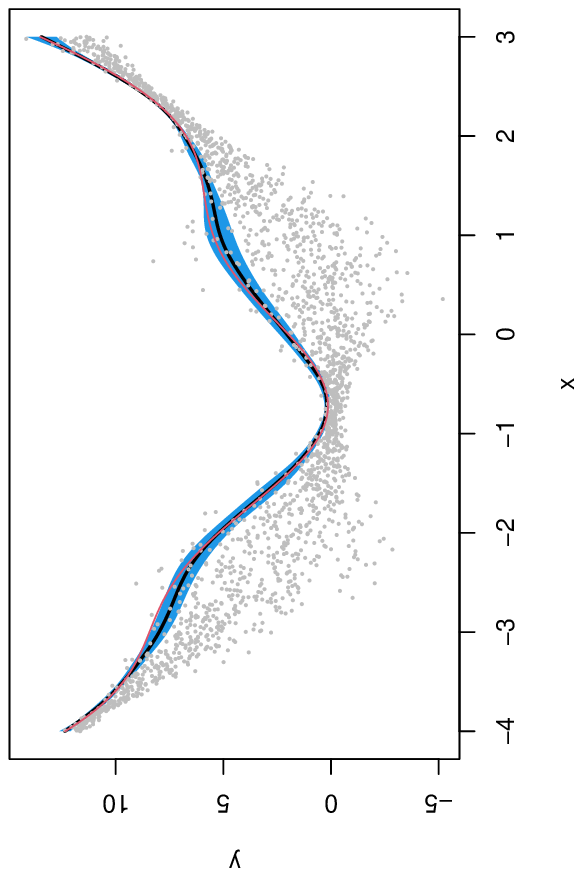}
\caption{Typical example NCV based smooth estimate of 95\% quantile for the distribution of the data shown in grey as a smooth function of $x$. The smooth is represented using a rank 50 penalized spline and the ELF loss is used. The fit is shown in black, with the 95\% credible band in blue and the truth in red.}\label{quantile-sim}
\end{figure}

\noindent {\bf Quantile regression example}. Smooth quantile regression models provide an example where attempts to use generalized cross validation perform exceptionally poorly. Moving away from the 50\% quantile, increasingly severe overfit is observed, as discussed in \cite{reiss.quantile}. The difficulty appears to relate to the averaging of leverage used in generalized cross validation, which is inappropriate for a highly asymmetric loss. The NCV approach avoids this problem and gives well calibrated inference. For example, the model $y \sim N(1+x+x^2,\{1.2+\sin(2x)\}^2)$ was used to simulate $x, y$ data for 2000 $x$ data evenly spaced over $[-4,3]$: see figure \ref{quantile-sim}. Smooth quantile regression was used to estimate the 95\% quantile, with the quantile represented using a rank 50 penalized cubic spline. The ELF loss of \cite{fasiolo2021qgam} was used as detailed in appendix \ref{sec:aqr}. The smoothing parameter was selected by leave-one-out NCV, and the coefficient covariance matrix computed using the method in section 
\ref{sec.uq}. 200 replicates were run, with a typical reconstruction and 95\% confidence band shown in figure \ref{quantile-sim}. Over 200 replicates the average across the function coverage probability of the nominally 95\% credible bands was 0.947, with on average 4.82\% of data lying above the estimated quantile. In this case the NCV criterion corresponds to the direct cross validation used by \cite{oh2011fast} (to within $O(p^3n^{-2})$ error), but they did not consider interval estimation and used direct calculation and grid search for the smoothing parameter optimization, which would be infeasibly expensive for multiple smoothing parameters, unlike NCV.

\section{Examples \label{sec.eg}}

\subsection{UK electricity load}

\begin{figure}
\eps{-90}{.4}{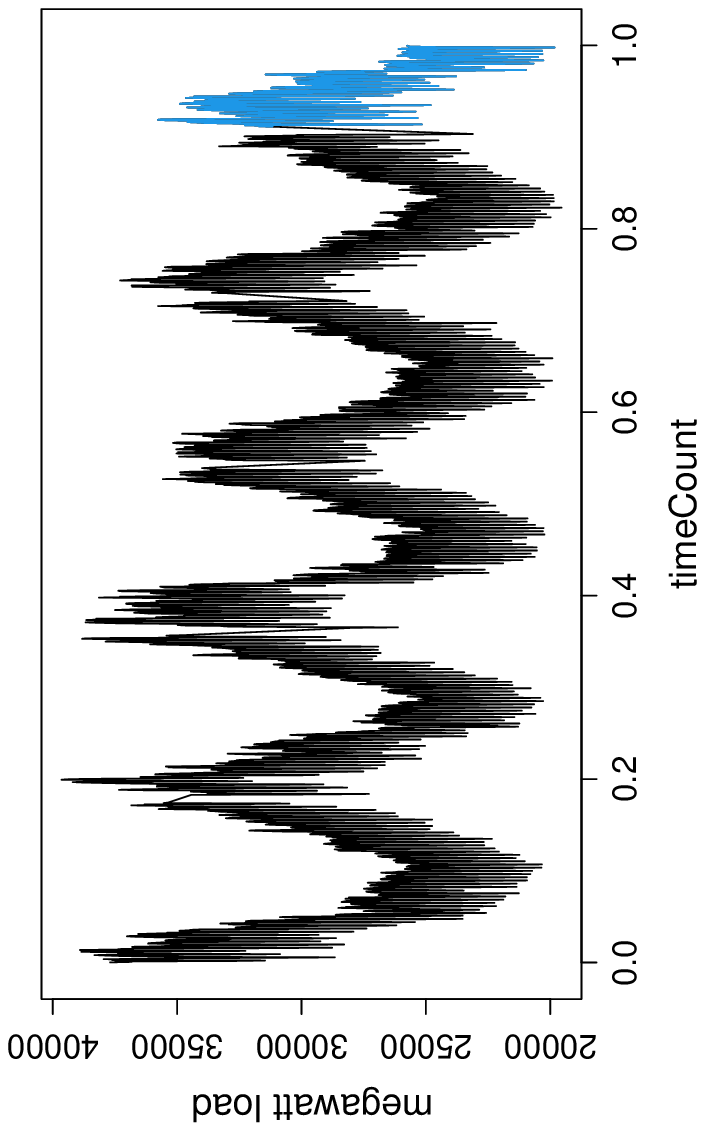}
\caption{Megawatt load on the UK electricity grid from 6AM until 6.30AM each day from 2011 to mid 2016, plotted against scaled time ({\tt timeCount}). The 2016 data shown in blue is to be predicted 24 hours ahead, from a model fitted to the data shown in black. }\label{UKL-data}
\end{figure}

Figure \ref{UKL-data} shows daily data on load on the UK national electricity grid between 6AM and 6.30AM\footnote{obtained from the predecessor of {\tt https://data. nationalgrideso.com/data-groups/demand} and {\tt https://demandforecast.nationalgrid.com/ efs\_demand\_forecast/faces/DataExplorer}}. Operational prediction of the load one day ahead is often based on generalized additive models, either alone or as the major component of a mixture of experts approach. There is typically a degree of un-modelled autocorrelation in the residuals from such models. This can cause over fit, increasing prediction error and degrading estimates of predictive uncertainty. A typical model structure for the data shown is 
\begin{multline*}
\E ({\tt load}_i) = \alpha_{{\tt dow}(i)} + f_{\tt dt(i)} ({\tt load48}_i) + f_1({\tt toy}_i) \\ + f_2({\tt timeCount}_i) + f_3({\tt temp}_i) + f_4({\tt temp95}_i)
\end{multline*}
where ${\tt dow}(i)$ is the day of the week of observation $i$ and  $\alpha_{{\tt dow}(i)}$ a parameter. The $f_j$ are all unknown smooth functions. $\tt dt(i)$ is a label taking one of 4 values: Mon, Sat, Sun or `ww' corresponding to any working day that follows a working day. ${\tt load48}_i$ is the load 48 half hours before point ${\tt load}_i$ occurred. So the idea is that load is predicted by the load at the same time the previous day, but that the relationship varies from day to day. ${\tt toy}_i$ is time of year, ${\tt timeCount}_i$ is scaled total elapsed time, ${\tt temp}_i$ is average current temperature and $ {\tt temp95}_i$ is exponentially smoothed lagged temperature. 

\begin{figure*}
\eps{-90}{.5}{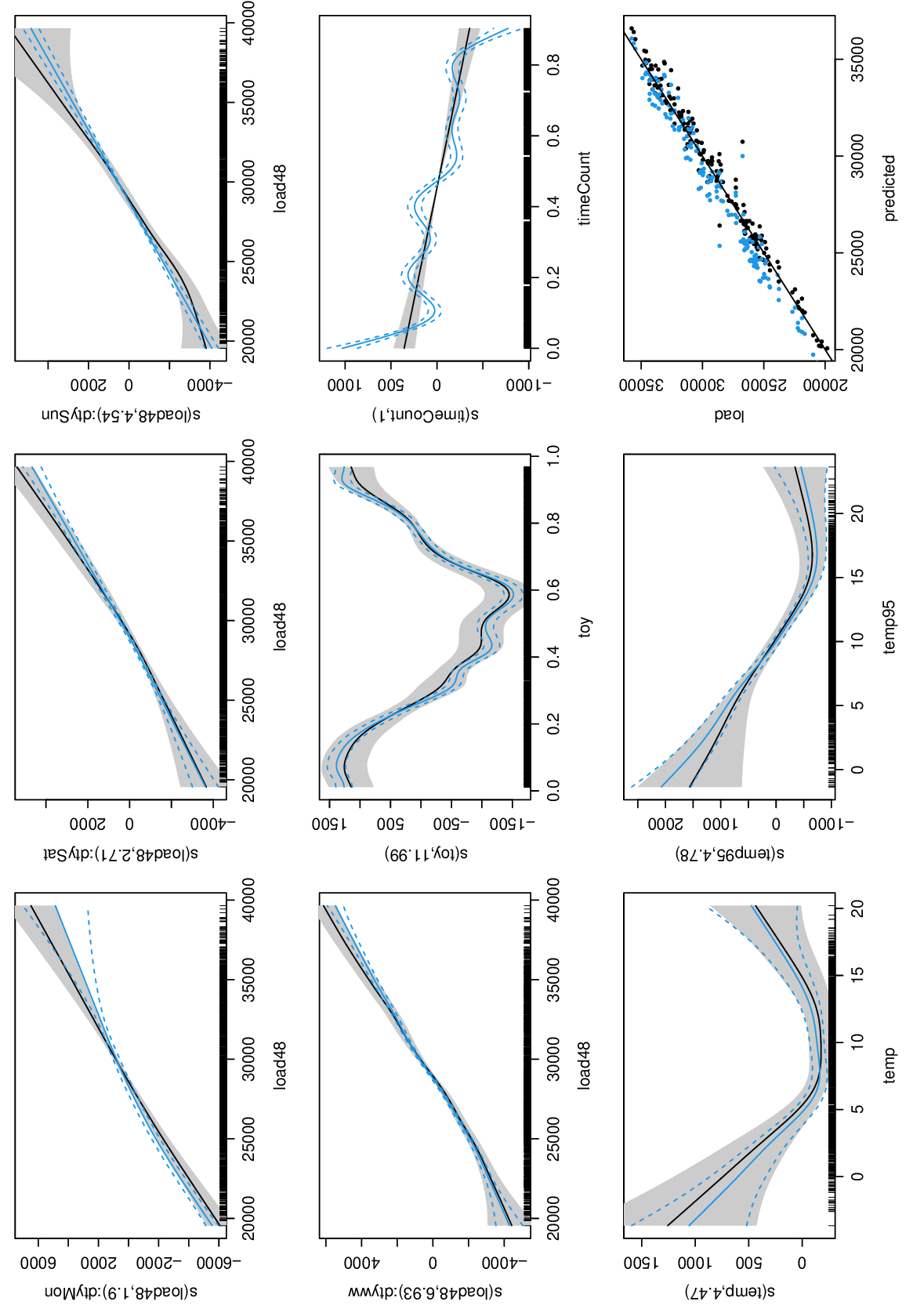}
\caption{Estimated smooth effects for the grid load prediction models. For the first 8 plots from top left, the black curves and grey bands show estimates and 95\% confidence bands from the NCV based model fit, while the superimposed blue curves show the equivalent for the marginal likelihood based fit, ignoring autocorrelation. The numbers in the y-axis labels are effective degrees of freedom. The plot at bottom right compares the models' predictive performance for the 2016 data. The black points show load versus predicted load for the NCV estimated model, and the blue points are for the marginal likelihood estimated model. The line illustrates where perfect prediction would lie.  
}\label{UKL-fit}
\end{figure*}

Fitting the model to pre-2016 data, assuming load is normally distributed with constant variance, and estimating smoothing parameters by standard marginal likelihood maximization gives significant residual autocorrelation to lag 9. Refitting the model by penalized least squares with NCV using the 9 neighbouring points on either side of point $i$ as $\alpha(i)$, results in smoother estimates for some model terms, as shown in figure \ref{UKL-fit}. It also yields improved prediction of load over 2016. Both models have mean absolute percentage prediction error (MAPE) of 1.1\% on the fit data. For prediction of 2016 this increases to 3.0\% for the standard approach, but only 1.6\% for the model estimated using NCV. Nominally 95\% prediction intervals have realized coverage of 65\% for the standard approach, and 91\% for the NCV based model. 

In this case it is clear that the improved prediction performance relates to the smoother {\tt timeCount} and {\tt toy} effects. Careful model construction could have imposed this in an ad hoc manner, and achieved MAPE performance similar to the NCV fit, but confidence bands would still be neglecting the residual autocorrelation. In addition operational use of these models typically involves substantial automation. For example, a full day's prediction would involve 48 models of the type shown here, which may in turn only be components of mixtures of experts, and which need to be regularly re-estimated as new training data accumulate. More local forecasting may involve thousands of such models. Hence the scope for careful ad hoc adjustment to deal with autocorrelation problems is limited, while the computational and implementational burden of fully specifying models for the autocorrelation appears prohibitive.  

An appealing alternative, when focused on forecasting, is to use a forecasting version of NCV. For example, one might predict for each week of the final year of fit data, so that for week $k$ of the final year $\alpha(k)$ would index all data from the start of week $k$ (or perhaps from the start of week $k-1$ onwards), while $\delta(k)$ would be the indices of the data for week $k$. Using this criterion produces effect estimates slightly smoother than those shown but practically indistinguishable MAPE results on fit and test data.        

\subsection{Swiss extreme rainfall}
\begin{figure*}
\eps{-90}{.5}{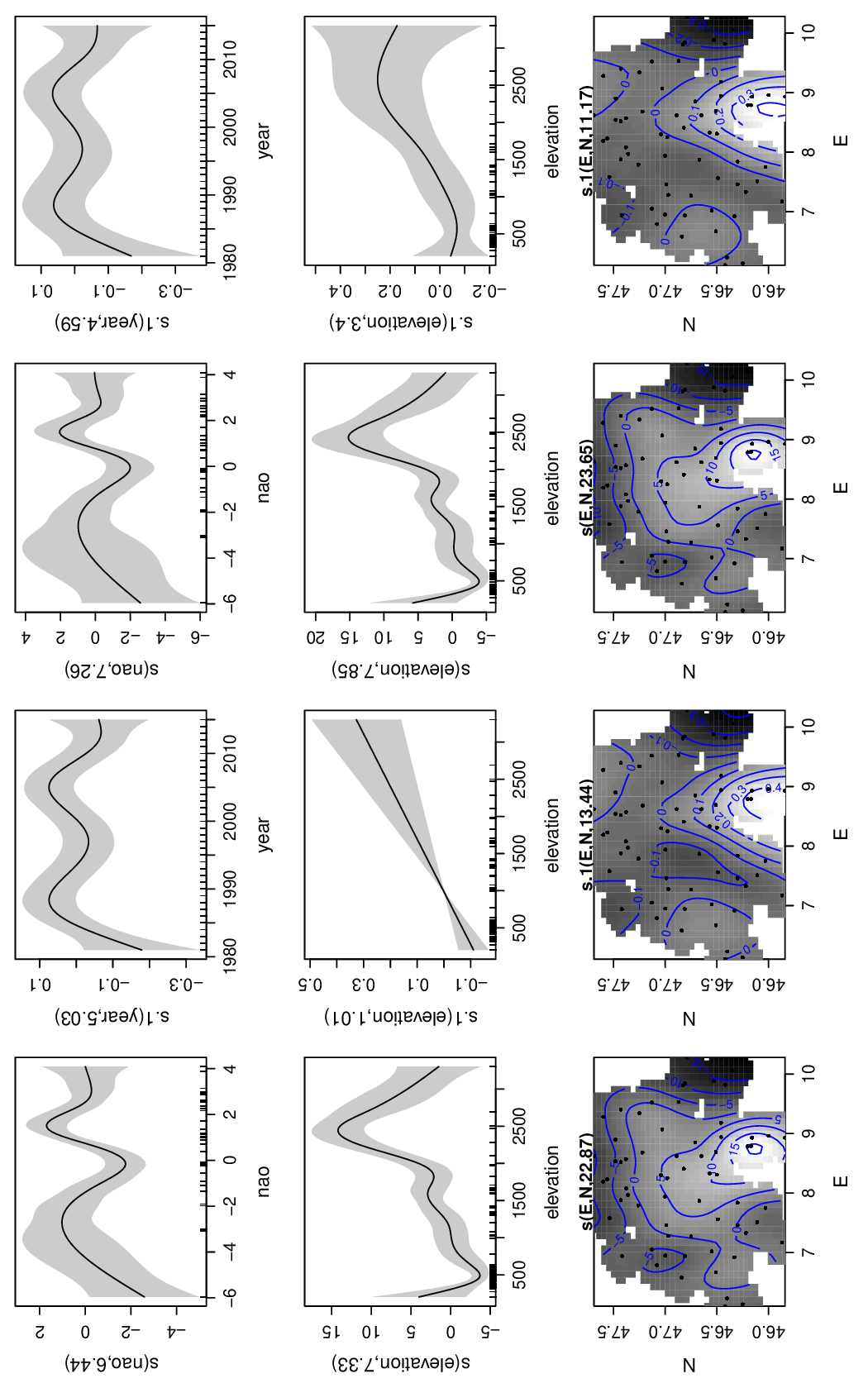}
\caption{Swiss extreme rainfall location scale model. The left two columns are for Laplace approximate marginal likelihood based smoothing parameter estimation, the right two for NCV with neighbourhoods based on stations closer than 0.3 degrees. In each column pair the left column shows smooth effect estimates for the location parameter, and the right for the log scale parameter. In this case the estimates are very similar between the methods, with little evidence for un-modelled short range spatial correlation being important here. A fit using leave-one-out NCV is visually almost indistinguishable from the marginal likelihood estimates.
}\label{swer}
\end{figure*}
Now considering a smooth location-scale model for the most extreme 12 hour total rainfall each year from 65 Swiss weather stations from 1981 to 2015\footnote{data are available as dataset {\tt swer} in R package {\tt gamair}.}. After some model selection \cite[][\S7.9.1]{wood2017igam}, the data were modelled using a generalized extreme value distribution with location parameter model $\mu = f_1(E,N) + f_2({\tt nao}) + f_3({\tt elevation}) + \gamma_{\tt c.r}$; log scale parameter model $\rho = \log \sigma = f_4(E,N) + f_5({\tt year}) + f_6({\tt elevation}) + \beta_{\tt c.r}$ and modified logit shape parameter model  $g(\xi) = \alpha_{\tt c.r}$. Function $g$ is a logit link modified to constrain $\xi$ between -1 and .5 (so variance remains finite and maximum likelihood estimation is consistent). The $f_j$ are smooth functions, represented using thin plate regression splines. There is one parameter $\alpha_j$, $\beta_j$ and $\gamma_j$ for each level of the 12 level factor ${\tt c.r}$, the climate region. Variable ${\tt nao}$ is the north Atlantic oscillation index. ${\tt N}$ and ${\tt E}$ are northing and easting in degrees. The station locations are shown as black points in the lower row of figure \ref{swer}.

The model can be estimated using the methods of \cite{wood2015plig}, with smoothing parameters estimated by Laplace approximate marginal likelihood. The resulting fit is shown in the left 2 columns of figure \ref{swer}. However, estimation of GEV models is somewhat statistically and numerically taxing, and it is hence useful to be able to check results by re-estimating with an alternative smoothing parameter estimation criterion. Leave one out NCV offers this possibility, using the variant of section \ref{qncv}, and in this case give results visually almost identical to those obtained via marginal likelihood. An additional concern for these data is the possibility of localised short range spatial autocorrelation in the extremes, with the accompanying danger of over-fit. To investigate this, NCV was used with neighbourhoods defined by stations being within 0.3 degrees of each other and year being the same (giving neighbourhood sizes from 1 to 6). The right 2 columns of figure  \ref{swer} show the results, which are little changed, suggesting little effect of short range spatial correlation here. So in this case the new methods allow considerably more confidence about the practical reliability of the results than was previously possible.

\subsection{Spatio-temporal models of forest health} 

\begin{figure*}
\eps{-90}{.5}{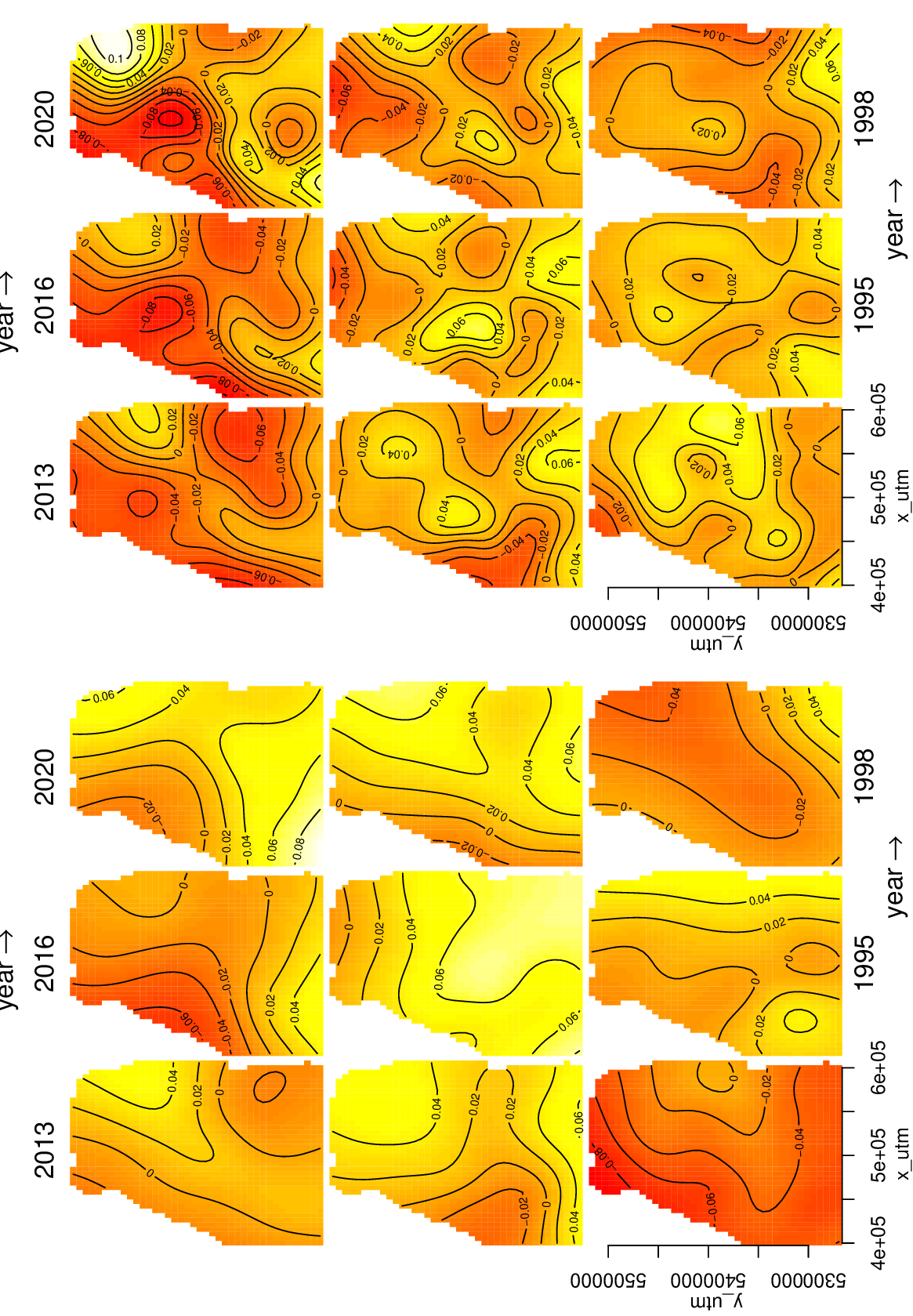}
\caption{Forest health model space-time interaction effect estimates. Left uses NCV accounting for the possibility of spatial and temporal autocorrelation as described in the text, and right uses REML assuming independence. Broadly, the NCV estimate is smoother in space, but less so in time. 
}\label{forest}
\end{figure*}

The final example concerns spatio-temporal modelling of forest health data for Norway spruce trees in Baden-W\"urttemberg, Germany. The data ($n=5876$) were gathered annually from 1991-2020 on a semi-regular 4km $\times$ 4km spatial grid, but with coarser (8km $\times$ 8km or 16km $\times$ 16km) sub-grids used in some years. The nodes of the grid are tree stands, and the response variable of interest is average defoliation of trees at the stand, as a proportion between 0 and 1 (average number of spruce trees per stand is 17). The survey is in alignment with the International Co-operative Programme on Assessment and Monitoring of Air Pollution Effects on Forests and thus uses the same survey protocol \citep{Eich2016}. There are reasons to expect some short range temporal autocorrelation at the stand level, as spruce do not shed all their needles each year. Within year local spatial correlation might also be present. Simply neglecting this, and treating the data as independent, risks over-fit and over-selection of model covariates. Neighbourhoods $\alpha(i)$ were hence constructed for each data point, consisting of the data from the same year and from a stand 10km or less away (on average 4 such neighbours per point), and from data less than 7 years apart at the same stand (average 5.6 such neighbours per point). The corresponding neighbourhood sizes range from 1 to 31 with a mean of 11.2. 

\begin{figure*}
\eps{-90}{.5}{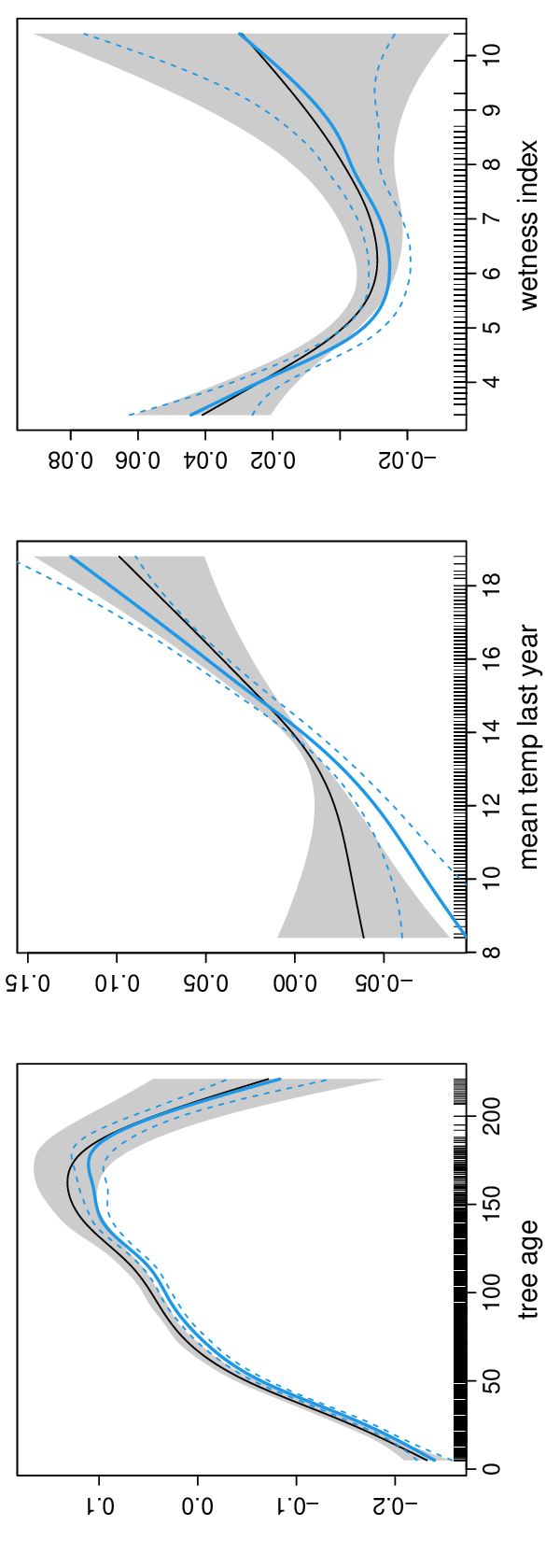}
\caption{The three most interesting smooth covariate effects for forest defoliation estimated by NCV (black and grey band) compared to REML assuming independence (blue). 
}\label{foreff}
\end{figure*}

Alongside spatial location and stand age (these are managed forests, so surveyed trees within a stand have very similar ages), a number of other covariates were measured, relating to the physical nature of the site (slope, elevation, aspect, wetness, soil composition, underlying geology, etc) and the tree properties (crown closure, level of fructification, forest diversity etc). Preliminary screening suggests some 16 covariates potentially of real interest (in addition to year and location). Generically the model for these data has the form 
$$
\E({\tt defoliation}_i) = f_0({\tt km.n}_i,{\tt km.e}_i,{\tt year}_i) + \sum_{j=1}^k f_j(x_{ji})
$$
where $x_j$ is a covariate and the $f_j$ are smooth functions. $f_0$ is a smooth space-time interaction, represented by a tensor product of a thin plate spline of space with a cubic spline of year \citep{augustin2009,eickenscheidt2018spatio}. The model can be estimated by penalized least squares with NCV smoothing parameter estimation, and a basic backward selection strategy used, which sequentially removes the least significant terms \citep[using the approximate p-values of][]{wood2013sp} until all are significant at the 10\% level. This approach retains 10 covariates: tree age, previous year's mean temperature, %tmean_veg_lag1
precipitation evaporation index in May, %spei_3_may
topographic wetness index, % twi25
topographic position index, %tpi500
exchangeably bound soil cations, %svals_molm2
days with max temperature $>25\text{C}$ %d_tmax25
and 3 others that barely reach the inclusion threshold. By contrast, repeating the same exercise, but neglecting the possibility of autocorrelation and using REML smoothness selection, retains 13 covariates, with the May precipitation/evaporation index replaced by the same for September, and a measure of forest diversity, %H_spec
net evapotranspiration the preceding year, % et0_veg_dv
and summed nitrate over critical load. 

Figure \ref{forest} compares the estimated spatio-temporal effects using NCV to those when the possibility of residual autocorrelation is ignored. The NCV estimates are smoother in space, but slightly less so in time. Figure \ref{foreff} picks out the three most interesting smooth covariate effect estimates and compares them when estimation is by NCV and REML assuming independence. There is slight attenuation of the time related effects, whereas the topographic wetness index effect appears stable.  The NCV model has about 20 fewer effective degrees of freedom, and an $r^2$ about 1.4\% lower than the alternative at just over 0.59. Clearly in this case there is evidence for short range residual autocorrelation in the data, as would be expected on a priori grounds, and this has some effect both on model estimates and the selection of model covariates.   
   
\section{Conclusions}

The NCV approach discussed here is both simple conceptually and general in applicability. It offers a principled method for choosing smoothing/precision parameters for models estimated with a variety of loss functions, in addition to negative log likelihoods and pseudo-likelihoods, and is also a model selection criterion \citep{arlot2010CV}. In the case in which the the loss is a sum of negative log likelihood components for each datum it is applicable in all situations for which Laplace approximate marginal likelihood (LAML) might otherwise be used \cite[see e.g.][]{wood2015plig}, but with the advantage over LAML of being directly useful for comparing models with differing fixed effects (unpenalized) structures. In any case, the availability of a second criterion for smoothing parameter selection in these cases offers a useful check of statistical stability in applications. The link to Jackknife estimation also leads to useful variance estimates. 

One immediate application is to the estimation of smooth models in the presence of nuisance autocorrelation. Addressing short range autocorrelation, without requiring formulation and estimation of a high rank correlation model, can offer some application advantages, potentially allowing fuller exploration of the model elements of most direct scientific interest. NCV always offers a way to check the sensitivity of model conclusions to the assumption of no un-modelled autocorrelation. However when such autocorrelation is present, it only offers mitigation when there is reason to expect that there is a separation of scales between the nuisance autocorrelation and the model components of interest. Otherwise there is no alternative but to fully model the correlation process. When the separation of scales assumption is reasonable, the question of whether better calibrated finite sample uncertainty quantification is possible for the NCV approach is interesting and open. A further open question is the feasibility of a general treatment of smoothing bias without requiring a Bayesian bias estimate via coherent belief updating.

The second immediate application is to non-likelihood based smooth loss functions. For example, in the smooth quantile regression case the approach offers a frequentist alternative to \cite{fasiolo2021qgam} without the need to subscribe to the Bayesian belief updating framework of \cite{bissiri16}, except for the limited purposes of accounting for smoothing bias in further inference. This less Bayesian approach aligns more readily with quantile regression's avoidance of a full probabilistic model. 

While the computational cost of NCV is the $O(np^2)$ of marginal likelihood or GCV, the constant of proportionality is higher and a disadvantage is that it is not obvious how to achieve the sort of scalability achieved for marginal likelihood by \cite{wood2016giga} and \cite{Li2019XWX} in big data settings. 

NCV is available in R package {\tt mgcv}. %Replication code is supplied as supplementary material \citep{wood2025sup}.

\appendix

\section{Quantile regression \label{sec:aqr}}

Following \cite{fasiolo2021qgam} the ELF loss targeting the $\tau$ quantile, $q$, is 
$$
\rho(y-\mu) = (\tau-1)\frac{y-q}{\sigma} + \lambda \log \left (1 + e^{\frac{y-q}{\lambda\sigma}} \right).
$$
This can be viewed as a smoothed version of the usual quantile regression pinball loss \citep{koenker2005}. As \cite{kaplan2017} show, smoothing the loss generally results in lower MSE than using the unsmoothed loss, and the degree of smoothing can be chosen to optimize MSE, with the optimum depending on the number of data per model degree of freedom. A simple approach is to fit a pilot location scale model to the data, to obtain standardized residuals $z_i = (y_i - \hat \mu_i)/\hat \sigma_i$: let $k$ be the estimated degrees of freedom of the pilot mean model. Taking size $\lceil n/k\rceil$ bootstrap samples, the MSE of $\hat q$ based on optimization of the ELF loss for each sample can be computed, where the $\tau$ quantile of the $z_i$ is the target truth. The $\lambda$ minimizing this MSE is the estimate of the optimal degree of loss smoothing, and is used for fitting the full model, along with the $\hat \sigma_i$ from the pilot fit. 

The approach of a pilot location-scale model fit and then computing the optimal amount of smoothing from the pooled pilot residuals follows \cite{fasiolo2021qgam}, but the bootstrap step simplifies the method for optimizing the smoothing of the loss, which is otherwise reliant on asymptotic results. Computationally, the same $n_b$ bootstrap resamples are used with each trial $\lambda$. This ensures that the MSE is smooth in $\lambda$. The optimal quantiles for each replicate can be found most efficiently by Newton optimization, starting from the true target quantile. The pilot fit consisted of a smooth model for the mean, using NCV, and then fitting a second smooth model to the resulting squared residuals to estimate $\hat \sigma^2_i$, again using NCV. A location scale model such as the {\tt gaulss} model in R package {\tt mgcv} would be an alternative. When using this smoothed loss, an obvious check is that $\sum_i \mathbb{I}(y_i<q_i) \approx \tau$.

\section{Example likelihood derivatives \label{sec:deriv.eg}}

This appendix supplies examples of the loss specific derivatives required in practice to compute the NCV derivatives as described in section \ref{sec:dncv}. First consider single parameter exponential family regression where $\mu_i = \E(y_i)$, $g(\mu_i) = \eta_i$, ${\bm \eta} = {\bf X}\bp$ and $\text{var}(y_i) = V(\mu_i)\phi$, for link function $g$, variance function $V$ and scale parameter $\phi$ (the loss is the negative log likelihood or the deviance). Then for, $l_i$, the contribution to the log likelihood from $y_i$ we have
$$
\pdif{l_i}{\beta_j} = \frac{y_i - \mu_i}{\phi g^\prime(\mu_i)V(\mu_i)} X_{ij}$$  and $$ \pddif{l_i}{\beta_j}{\rho_q} = - \frac{X_{ij}\alpha(\mu_i)}{g^\prime(\mu_i)^2V(\mu_i) } \sum_k X_{ik} \dif{\hat \beta_k}{\rho_q}.
$$
The latter term  provides $\ildif{{\bf g}}{\rho_j}$. More generally for other location parameter regression cases,
$$
\pdif{l_i}{\beta_j} = \pdif{l_i}{\mu_i}\dif{\mu_i}{\eta_i} X_{ij} $$ and $$ \pddif{l_i}{\beta_j}{\rho_q} =
\left \{  \pdif{^2 l_i}{\mu_i^2} \left ( \dif{\mu_i}{\eta_i} \right )^2 + \pdif{l_i}{\mu_i} \dif{^2 \mu_i}{\eta_i^2}
\right \} X_{ij} \sum_k X_{ik} \dif{\hat \beta_k}{\rho_q}.
$$
In both these cases the NCV criterion has the form 
$
V = \sum_{k=1}^m\sum_{i \in \beta(k)}{\cal D}(y_i,\eta_i^{-\alpha(k)})
$
so that 
$$
\pdif{V}{\rho_j} = \sum_{k=1}^m\sum_{i \in \beta(k)} \left .\pdif{\cal D}{\eta_i}\right |_{{\eta}_i^{-\alpha(i)}} \pdif{{\eta}_i^{-\alpha(i)}}{\rho_j}
$$
Some location parameter regressions may also depend on extra parameters, $\bm \theta$ (not varying with $i$). Then $ V = \sum_{k=1}^m\sum_{i \in \beta(k)} {\cal D}(y_i,{\eta}_i^{-\alpha(i)},{\bm \theta}) $ so that
$$
\pdif{V}{\theta_j} = \sum_{k=1}^m\sum_{i \in \beta(k)} \left .\pdif{\cal D}{\eta_i}\right |_{{\eta}_i^{-\alpha(i)}} \pdif{{\eta}_i^{-\alpha(i)}}{\theta_j} + \left .\pdif{\cal D}{\theta_j}\right |_{{\eta}_i^{-\alpha(i)}}
$$
and
\begin{multline*}
\pddif{l_i}{\beta_j}{\theta_q} =
\left \{  \pdif{^2 l_i}{\mu_i^2} \left ( \dif{\mu_i}{\eta_i} \right )^2 + \pdif{l_i}{\mu_i} \dif{^2 \mu_i}{\eta_i^2}
\right \} X_{ij} \sum_k X_{ik} \dif{\hat \beta_k}{\theta_q} \\+ \pddif{l_i}{\mu_i}{\theta_q} \dif{\mu_i}{\eta_i} X_{ij}.
\end{multline*}
The Hessian and its derivative w.r.t. $\rho_j$ follow in a similar manner.

\section{Low level computational considerations \label{sec:low-level}}

The $O(np^2)$ floating point operation cost of NCV is similar to the cost for GCV or REML. However in reality NCV is more costly in practice, since its leading order operations consist of matrix-vector operations, while GCV and REML computations can be structured so that the leading order operations are matrix-matrix. Matrix-matrix operations can typically be arranged in a way that is highly cache efficient and any optimized BLAS \citep[e.g.][]{openblas} will exploit this. At time of writing such cache efficiency typically results in some 20-fold speed up for matrix cross products, for example. Matrix-vector operations can not be arranged in this cache efficient manner \cite[see][for an introduction to the issues]{GvL4}. 

However, this disadvantage can be offset by relatively simple parallelization. The terms in the summations (\ref{NCV}) can be trivially parallelized, as can the summations required for derivative computation. Here, OMP \citep{openmp08} in C was used. Realized scaling is then reasonable. This is in contrast to attempts to use simple parallelization approaches for matrix-matrix dominated GCV or REML computations. In such cases anything gained by parallelization is often immediately lost, as different execution threads compete for cache, destroying the cache efficiency on which the BLAS relies. For matrix-vector dominated computation this trade-off does not occur. 

\section{variance estimation under penalization\label{sec:pen-var}}

In a penalised regression setting the convergence rates for $\hat {\bf V}$, from section \ref{sec.uq}, will be reduced, since the best that we can then do is $\hat {\bm \delta} = O_p(n^{-4/9})$ \citep[e.g.][]{claeskens2009}. As a concrete example, for a one dimensional cubic regression spline sieve estimator with $k$ knots we could set $k \propto n^{1/9}$ \citep{kauermann2009}. Again consider reconstruction of a fixed function under infill asymptotics, where none the less $|\alpha(i)|$ does not grow with $n$, so that a separation of scales occurs between the autocorrelated noise process and the signal. Then $b_{ki} = O(n^{-8/9})$ and we now have
\begin{align*}
n^{8/9}(\hat V_{km} -V_{km}) &= n \sum_i^n b_{ki}\sum_{j \in \alpha(i)} b_{mj} \{e_ie_j - \E(e_ie_j)\}\\ &= O_p(n^{-7/18}),
\end{align*} 
a somewhat reduced rate. Considering the bias terms caused by the correlation of $\hat \delta_i$ and the residuals indexed by $\alpha(i)$ we now have
$$
n^{8/9}\sum b_{ki} (y_i - \hat \mu_i) \sum_{j\in \alpha(i)}b_{mj} \hat \delta_j = O_p(n^{-1/3})
$$
a lowered convergence rate (this bias is no longer simply increasing the error at any given $n$ without changing the rate). Again if the residuals are independent of $\hat \delta_i$ then these bias terms are $O_p(n^{-5/6})$ and therefore negligible. 

Hence as in the main paper, provided we use cross-validated residuals in place of the usual residuals, the estimator of the covariance matrix can be expected to be reasonable, as long as there is really a separation of scales between the signal and the autocorrelation scale length. 

If the $\sqrt{n}$ scaling of the random walk appears non-obvious because of the induced correlation, note that we could decompose the summations into $\max |\alpha(i)|$ separate random walks each with $O(n)$ finite variance independent increments. Each of these components is clearly $O_p(n^{1/2})$ as is the sum of a fixed finite number of them.    

\stas
\begin{acks}[Acknowledgments]
I thank Heike Puhlmann and Simon Trust at the Forest Research Institute Baden-W\"urttemberg, Germany for the Terrestrial Crown Condition Inventory (TCCI) forest health monitoring survey data, and Nicole Augustin for the corresponding model structure. Thanks to the reviewers for comments on improving the clarity of the paper.
\end{acks}

\begin{supplement}
\stitle{Supporting code and data}
\sdescription{Code and data for replication of the results in sections \ref{sec.sim} and \ref{sec.eg}  are provided in the supplementary material \citep{wood2025sup}.}
\end{supplement}

\else
\subsection*{Acknowledgments}
I thank Heike Puhlmann and Simon Trust at the Forest Research Institute Baden-W\"urttemberg, Germany for the Terrestrial Crown Condition Inventory (TCCI) forest health monitoring survey data, and Nicole Augustin for the corresponding model structure. Thanks to the reviewers for comments on improving the clarity of the paper.
\fi
\stas
%\bibliography{/home/sw283/bibliography/journal,/home/sw283/bibliography/simon}

\begin{thebibliography}{}

\bibitem[\protect\citeauthoryear{Arlot and Celisse}{Arlot and
  Celisse}{2010}]{arlot2010CV}
Arlot, S. and A.~Celisse (2010).
\newblock A survey of cross-validation procedures for model selection.
\newblock {\em Statistics Surveys\/}~{\em 4}, 40--79.

\bibitem[\protect\citeauthoryear{Augustin, Musio, von Wilpert, Kublin, Wood,
  and Schumacher}{Augustin et~al.}{2009}]{augustin2009}
Augustin, N.~H., M.~Musio, K.~von Wilpert, E.~Kublin, S.~N. Wood, and
  M.~Schumacher (2009).
\newblock Modeling spatiotemporal forest health monitoring data.
\newblock {\em Journal of the American Statistical Association\/}~{\em
  104\/}(487), 899--911.

\bibitem[\protect\citeauthoryear{Bates, Hastie, and Tibshirani}{Bates
  et~al.}{2024}]{bates2024cv}
Bates, S., T.~Hastie, and R.~Tibshirani (2024).
\newblock Cross-validation: what does it estimate and how well does it do it?
\newblock {\em Journal of the American Statistical Association\/}~{\em
  119\/}(546), 1434--1445.

\bibitem[\protect\citeauthoryear{Beirami, Razaviyayn, Shahrampour, and
  Tarokh}{Beirami et~al.}{2017}]{beirami2017approxcv}
Beirami, A., M.~Razaviyayn, S.~Shahrampour, and V.~Tarokh (2017).
\newblock On optimal generalizability in parametric learning.
\newblock {\em Advances in neural information processing systems\/}~{\em 30}.

\bibitem[\protect\citeauthoryear{Bissiri, Holmes, and Walker}{Bissiri
  et~al.}{2016}]{bissiri16}
Bissiri, P.~G., C.~Holmes, and S.~G. Walker (2016).
\newblock A general framework for updating belief distributions.
\newblock {\em Journal of the Royal Statistical Society. Series B\/}~{\em
  78\/}(5), 1103--1130.

\bibitem[\protect\citeauthoryear{Chu and Marron}{Chu and
  Marron}{1991}]{chu1991MCV}
Chu, C.-K. and J.~S. Marron (1991).
\newblock Comparison of two bandwidth selectors with dependent errors.
\newblock {\em The Annals of Statistics\/}~{\em 19\/}(4), 1906--1918.

\bibitem[\protect\citeauthoryear{Claeskens, Krivobokova, and Opsomer}{Claeskens
  et~al.}{2009}]{claeskens2009}
Claeskens, G., T.~Krivobokova, and J.~D. Opsomer (2009).
\newblock Asymptotic properties of penalized spline estimators.
\newblock {\em Biometrika\/}~{\em 96\/}(3), 529--544.

\bibitem[\protect\citeauthoryear{Cox and Reid}{Cox and
  Reid}{2004}]{coxreid2004}
Cox, D.~R. and N.~Reid (2004).
\newblock A note on pseudolikelihood constructed from marginal densities.
\newblock {\em Biometrika\/}~{\em 91\/}(3), 729--737.

\bibitem[\protect\citeauthoryear{Craven and Wahba}{Craven and
  Wahba}{1979}]{craven.wahba}
Craven, P. and G.~Wahba (1979).
\newblock Smoothing noisy data with spline functions.
\newblock {\em Numerische Mathematik\/}~{\em 31\/}(5), 377--403.

\bibitem[\protect\citeauthoryear{Davison and Hinkley}{Davison and
  Hinkley}{1997}]{davison1997bootstrap}
Davison, A.~C. and D.~V. Hinkley (1997).
\newblock {\em Bootstrap methods and their application}.
\newblock Cambridge university press.

\bibitem[\protect\citeauthoryear{deHoog and Hutchinson}{deHoog and
  Hutchinson}{1987}]{deHoog1987}
deHoog, F. and M.~Hutchinson (1987).
\newblock An efficient method for calculating smoothing splines using
  orthogonal transformations.
\newblock {\em Numerische Mathematik\/}~{\em 50}, 311--319.

\bibitem[\protect\citeauthoryear{Eichhorn, Roskams, Poto\`ci\`c, Timmermann,
  Ferretti, Mues, Szepesi, Durrant, Seletkovi\'c, H-W.Schr\"ock, Nevalainen,
  Bussotti, Garcia, and Wulff}{Eichhorn et~al.}{2017}]{Eich2016}
Eichhorn, J., P.~Roskams, N.~Poto\`ci\`c, V.~Timmermann, Ferretti, V.~Mues,
  A.~Szepesi, D.~Durrant, I.~Seletkovi\'c, H-W.Schr\"ock, S.~Nevalainen,
  F.~Bussotti, P.~Garcia, and S.~Wulff (Eds.) (2017).
\newblock {\em {ICP} {F}orests manual on methods and criteria for harmonized
  sampling, assessment, monitoring and analysis of the effects of air pollution
  on forests.}
\newblock Th\"unen Institute of Forest Ecosystems, Eberswalde,Germany.

\bibitem[\protect\citeauthoryear{Eickenscheidt, Augustin, and
  Wellbrock}{Eickenscheidt et~al.}{2019}]{eickenscheidt2018spatio}
Eickenscheidt, N., N.~H. Augustin, and N.~Wellbrock (2019).
\newblock Spatio-temporal modelling of forest monitoring data: {M}odelling
  {G}erman tree defoliation data collected between 1989 and 2015 for trend
  estimation and survey grid examination using {GAMM}s.
\newblock {\em iForest Biogeosciences and Forestry\/}~{\em 12}, 338--348.

\bibitem[\protect\citeauthoryear{Eld{\'e}n}{Eld{\'e}n}{1984}]{elden84}
Eld{\'e}n, L. (1984).
\newblock A note on the computation of the generalized cross-validation
  function for ill-conditioned least squares problems.
\newblock {\em BIT Numerical Mathematics\/}~{\em 24\/}(4), 467--472.

\bibitem[\protect\citeauthoryear{Fasiolo, Wood, Zaffran, Nedellec, and
  Goude}{Fasiolo et~al.}{2021}]{fasiolo2021qgam}
Fasiolo, M., S.~N. Wood, M.~Zaffran, R.~Nedellec, and Y.~Goude (2021).
\newblock Fast calibrated additive quantile regression.
\newblock {\em Journal of the American Statistical Association\/}~{\em
  116\/}(535), 1402--1412.

\bibitem[\protect\citeauthoryear{Golub, Heath, and Wahba}{Golub
  et~al.}{1979}]{golub.heath.wahba}
Golub, G.~H., M.~Heath, and G.~Wahba (1979).
\newblock Generalized cross validation as a method for choosing a good ridge
  parameter.
\newblock {\em Technometrics\/}~{\em 21\/}(2), 215--223.

\bibitem[\protect\citeauthoryear{Golub and van Loan}{Golub and van
  Loan}{2013}]{GvL4}
Golub, G.~H. and C.~F. van Loan (2013).
\newblock {\em Matrix {C}omputations\/} (4th ed.).
\newblock Baltimore: Johns Hopkins University Press.

\bibitem[\protect\citeauthoryear{Gouri{\'e}roux, Monfort, and
  Renault}{Gouri{\'e}roux et~al.}{2017}]{gourieroux2017pseudolikelihood}
Gouri{\'e}roux, C., A.~Monfort, and E.~Renault (2017).
\newblock Consistent pseudo-maximum likelihood estimators.
\newblock {\em Annals of Economics and Statistics/Annales d'{\'E}conomie et de
  Statistique\/}~(125/126), 187--218.

\bibitem[\protect\citeauthoryear{Gu and Xiang}{Gu and Xiang}{2001}]{gu2001gacv}
Gu, C. and D.~Xiang (2001).
\newblock Cross-validating non-gaussian data: Generalized approximate
  cross-validation revisited.
\newblock {\em Journal of Computational and Graphical Statistics\/}~{\em
  10\/}(3), 581--591.

\bibitem[\protect\citeauthoryear{Harville}{Harville}{1997}]{harville:1997}
Harville, D.~A. (1997).
\newblock {\em Matrix Algebra from a Statistician's Perspective}.
\newblock New York: Springer.

\bibitem[\protect\citeauthoryear{Hastie and Tibshirani}{Hastie and
  Tibshirani}{1990}]{h&t90}
Hastie, T. and R.~Tibshirani (1990).
\newblock {\em Generalized Additive Models}.
\newblock Chapman \& Hall.

\bibitem[\protect\citeauthoryear{Hastie, Tibshirani, and Friedman}{Hastie
  et~al.}{2009}]{ESL2009}
Hastie, T., R.~Tibshirani, and J.~Friedman (2009).
\newblock {\em The {E}lements of {S}tatistical {L}earning}.
\newblock Springer.

\bibitem[\protect\citeauthoryear{Hutchinson and De~Hoog}{Hutchinson and
  De~Hoog}{1985}]{hutchinson1985}
Hutchinson, M.~F. and F.~De~Hoog (1985).
\newblock Smoothing noisy data with spline functions.
\newblock {\em Numerische Mathematik\/}~{\em 47\/}(1), 99--106.

\bibitem[\protect\citeauthoryear{Kaplan and Sun}{Kaplan and
  Sun}{2017}]{kaplan2017}
Kaplan, D.~M. and Y.~Sun (2017).
\newblock Smoothed estimating equations for instrumental variables quantile
  regression.
\newblock {\em Econometric Theory\/}~{\em 33\/}(1), 105--157.

\bibitem[\protect\citeauthoryear{Kauermann, Krivobokova, and
  Fahrmeir}{Kauermann et~al.}{2009}]{kauermann2009}
Kauermann, G., T.~Krivobokova, and L.~Fahrmeir (2009).
\newblock Some asymptotic results on generalized penalized spline smoothing.
\newblock {\em Journal of the Royal Statistical Society: Series B (Statistical
  Methodology)\/}~{\em 71\/}(2), 487--503.

\bibitem[\protect\citeauthoryear{Klein, Kneib, Klasen, and Lang}{Klein
  et~al.}{2014}]{klein2014dr}
Klein, N., T.~Kneib, S.~Klasen, and S.~Lang (2014).
\newblock Bayesian structured additive distributional regression for
  multivariate responses.
\newblock {\em Journal of the Royal Statistical Society: Series C (Applied
  Statistics)\/}~{\em 64}, 569--591.

\bibitem[\protect\citeauthoryear{Klein, Kneib, Lang, and Sohn}{Klein
  et~al.}{2015}]{klein2015dr}
Klein, N., T.~Kneib, S.~Lang, and A.~Sohn (2015).
\newblock Bayesian structured additive distributional regression with an
  application to regional income inequality in {G}ermany.
\newblock {\em Annals of Applied Statistics\/}~{\em 9}, 1024--1052.

\bibitem[\protect\citeauthoryear{Koenker}{Koenker}{2005}]{koenker2005}
Koenker, R. (2005).
\newblock {\em Quantile Regression}.
\newblock Cambridge University Press.

\bibitem[\protect\citeauthoryear{Krivobokova and Kauermann}{Krivobokova and
  Kauermann}{2007}]{krivobokova2007}
Krivobokova, T. and G.~Kauermann (2007).
\newblock A note on penalized spline smoothing with correlated errors.
\newblock {\em Journal of the American Statistical Association\/}~{\em
  102\/}(480), 1328--1337.

\bibitem[\protect\citeauthoryear{Li and Wood}{Li and Wood}{2020}]{Li2019XWX}
Li, Z. and S.~N. Wood (2020).
\newblock Faster model matrix crossproducts for large generalized linear models
  with discretized covariates.
\newblock {\em Statistics and Computing.\/}~{\em 30\/}(1), 19--25.

\bibitem[\protect\citeauthoryear{Lukas}{Lukas}{2006}]{lukas2006robust}
Lukas, M.~A. (2006).
\newblock Robust generalized cross-validation for choosing the regularization
  parameter.
\newblock {\em Inverse Problems\/}~{\em 22\/}(5), 1883.

\bibitem[\protect\citeauthoryear{Lukas}{Lukas}{2010}]{lukas2010rgcv}
Lukas, M.~A. (2010).
\newblock Robust {GCV} choice of the regularization parameter for correlated
  data.
\newblock {\em The Journal of integral equations and applications\/}, 519--547.

\bibitem[\protect\citeauthoryear{Lukas, de~Hoog, and Anderssen}{Lukas
  et~al.}{2016}]{lukas2016}
Lukas, M.~A., F.~R. de~Hoog, and R.~S. Anderssen (2016).
\newblock Practical use of robust {GCV} and modified {GCV} for spline
  smoothing.
\newblock {\em Computational Statistics\/}~{\em 31\/}(1), 269--289.

\bibitem[\protect\citeauthoryear{Marra and Wood}{Marra and
  Wood}{2012}]{marra.wood2012}
Marra, G. and S.~N. Wood (2012).
\newblock Coverage properties of confidence intervals for generalized additive
  model components.
\newblock {\em Scandinavian Journal of Statistics\/}~{\em 39\/}(1), 53--74.

\bibitem[\protect\citeauthoryear{Mayr, Fenske, Hofner, Kneib, and Schmid}{Mayr
  et~al.}{2012}]{mayr2012gamlssboost}
Mayr, A., N.~Fenske, B.~Hofner, T.~Kneib, and M.~Schmid (2012).
\newblock Generalized additive models for location, scale and shape for high
  dimensional data --- a flexible approach based on boosting.
\newblock {\em Journal of the Royal Statistical Society. Series C\/}~{\em
  61\/}(3), 403--427.

\bibitem[\protect\citeauthoryear{Nocedal and Wright}{Nocedal and
  Wright}{2006}]{nocedal.wright}
Nocedal, J. and S.~Wright (2006).
\newblock {\em Numerical {O}ptimization\/} (2nd ed.).
\newblock New York: Springer {V}erlag.

\bibitem[\protect\citeauthoryear{Nychka}{Nychka}{1988}]{nychka88}
Nychka, D. (1988).
\newblock Bayesian confidence intervals for smoothing splines.
\newblock {\em Journal of the American Statistical Association\/}~{\em
  83\/}(404), 1134--1143.

\bibitem[\protect\citeauthoryear{Oh, Lee, and Nychka}{Oh
  et~al.}{2011}]{oh2011fast}
Oh, H.-S., T.~C. Lee, and D.~W. Nychka (2011).
\newblock Fast nonparametric quantile regression with arbitrary smoothing
  methods.
\newblock {\em Journal of Computational and Graphical Statistics\/}~{\em
  20\/}(2), 510--526.

\bibitem[\protect\citeauthoryear{{OpenMP Architecture Review Board}}{{OpenMP
  Architecture Review Board}}{2008}]{openmp08}
{OpenMP Architecture Review Board} (2008, May).
\newblock {OpenMP} application program interface version 3.0.

\bibitem[\protect\citeauthoryear{Opsomer, Wang, and Yang}{Opsomer
  et~al.}{2001}]{opsomer2001}
Opsomer, J., Y.~Wang, and Y.~Yang (2001).
\newblock Nonparametric regression with correlated errors.
\newblock {\em Statistical Science\/}, 134--153.

\bibitem[\protect\citeauthoryear{Paige and Saunders}{Paige and
  Saunders}{1975}]{paige1975symmlq}
Paige, C.~C. and M.~A. Saunders (1975).
\newblock Solution of sparse indefinite systems of linear equations.
\newblock {\em SIAM journal on numerical analysis\/}~{\em 12\/}(4), 617--629.

\bibitem[\protect\citeauthoryear{Rad and Maleki}{Rad and
  Maleki}{2020}]{rad2020approxcv}
Rad, K.~R. and A.~Maleki (2020).
\newblock A scalable estimate of the out-of-sample prediction error via
  approximate leave-one-out cross-validation.
\newblock {\em Journal of the Royal Statistical Society Series B: Statistical
  Methodology\/}~{\em 82\/}(4), 965--996.

\bibitem[\protect\citeauthoryear{Reiss and Huang}{Reiss and
  Huang}{2012}]{reiss.quantile}
Reiss, P.~T. and L.~Huang (2012).
\newblock Smoothness selection for penalized quantile regression splines.
\newblock {\em The international journal of biostatistics\/}~{\em 8\/}(1).

\bibitem[\protect\citeauthoryear{Rigby and Stasinopoulos}{Rigby and
  Stasinopoulos}{2005}]{gamlss}
Rigby, R. and D.~M. Stasinopoulos (2005).
\newblock Generalized additive models for location, scale and shape (with
  discussion).
\newblock {\em Journal of the Royal Statistical Society. Series C\/}~{\em
  54\/}(3), 507--554.

\bibitem[\protect\citeauthoryear{Roberts, Bahn, Ciuti, Boyce, Elith,
  Guillera-Arroita, Hauenstein, Lahoz-Monfort, Schr{\"o}der, Thuiller,
  et~al.}{Roberts et~al.}{2017}]{roberts2017CV}
Roberts, D.~R., V.~Bahn, S.~Ciuti, M.~S. Boyce, J.~Elith, G.~Guillera-Arroita,
  S.~Hauenstein, J.~J. Lahoz-Monfort, B.~Schr{\"o}der, W.~Thuiller, et~al.
  (2017).
\newblock Cross-validation strategies for data with temporal, spatial,
  hierarchical, or phylogenetic structure.
\newblock {\em Ecography\/}~{\em 40\/}(8), 913--929.

\bibitem[\protect\citeauthoryear{Robinson and Moyeed}{Robinson and
  Moyeed}{1989}]{robinson1989rcv}
Robinson, T. and R.~Moyeed (1989).
\newblock Making robust the cross-validatory choice of smoothing parameter in
  spline smoothing regression.
\newblock {\em Communications in Statistics-Theory and Methods\/}~{\em
  18\/}(2), 523--539.

\bibitem[\protect\citeauthoryear{Shi}{Shi}{1988}]{shi88}
Shi, X. (1988).
\newblock A note on the delete-d jackknife variance estimators.
\newblock {\em Statistics \& probability letters\/}~{\em 6\/}(5), 341--347.

\bibitem[\protect\citeauthoryear{Stasinopoulos, Rigby, Heller, Voudouris, and
  De~Bastiani}{Stasinopoulos et~al.}{2017}]{gamlss.book}
Stasinopoulos, M.~D., R.~A. Rigby, G.~Z. Heller, V.~Voudouris, and
  F.~De~Bastiani (2017).
\newblock {\em Flexible {R}egression and {S}moothing: using {GAMLSS} in {R}}.
\newblock Chapman and Hall/CRC.

\bibitem[\protect\citeauthoryear{Stephenson and Broderick}{Stephenson and
  Broderick}{2020}]{stephenson2020loocv}
Stephenson, W. and T.~Broderick (2020).
\newblock Approximate cross-validation in high dimensions with guarantees.
\newblock In {\em International Conference on Artificial Intelligence and
  Statistics}, pp.\  2424--2434. PMLR.

\bibitem[\protect\citeauthoryear{Stone}{Stone}{1974}]{stone74}
Stone, M. (1974).
\newblock Cross-validatory choice and assessment of statistical predictions
  (with discussion).
\newblock {\em Journal of the Royal Statistical Society, Series B\/}~{\em 36},
  111--147.

\bibitem[\protect\citeauthoryear{Stone}{Stone}{1977}]{stone77}
Stone, M. (1977).
\newblock An asymptotic equivalence of choice of model by cross-validation and
  {A}kaike's criterion.
\newblock {\em Journal of the Royal Statistical Society, Series B\/}~{\em 39},
  44--47.

\bibitem[\protect\citeauthoryear{van~der Linde}{van~der
  Linde}{2000}]{vanderLinde2000}
van~der Linde, A. (2000).
\newblock {\em Smoothing and {R}egression}, Chapter Variance {E}stimation and
  {S}moothing-{P}arameter {S}election for {S}pline {R}egression, pp.\  19--42.
\newblock Wiley.

\bibitem[\protect\citeauthoryear{van~der Vorst}{van~der
  Vorst}{2003}]{vandervorst2003}
van~der Vorst, H.~A. (2003).
\newblock {\em Iterative {K}rylov {M}ethods for {L}arge {L}inear {S}ystems}.
\newblock Cambridge University Press.

\bibitem[\protect\citeauthoryear{Wahba}{Wahba}{1985}]{wahba85}
Wahba, G. (1985).
\newblock A comparison of {GCV} and {GML} for choosing the smoothing parameter
  in the generalized spline smoothing problem.
\newblock {\em Annals of Statistics\/}~{\em 13\/}(4), 1378--1402.

\bibitem[\protect\citeauthoryear{Wilson, Kasy, and Mackey}{Wilson
  et~al.}{2020}]{wilson2020approxcv}
Wilson, A., M.~Kasy, and L.~Mackey (2020).
\newblock Approximate cross-validation: Guarantees for model assessment and
  selection.
\newblock In {\em International conference on artificial intelligence and
  statistics}, pp.\  4530--4540. PMLR.

\bibitem[\protect\citeauthoryear{Wood}{Wood}{2008}]{wood2008gacv}
Wood, S.~N. (2008).
\newblock Fast stable direct fitting and smoothness selection for generalized
  additive models.
\newblock {\em Journal of the Royal Statistical Society: Series B (Statistical
  Methodology)\/}~{\em 70\/}(3), 495--518.

\bibitem[\protect\citeauthoryear{Wood}{Wood}{2011}]{wood2011}
Wood, S.~N. (2011).
\newblock Fast stable restricted maximum likelihood and marginal likelihood
  estimation of semiparametric generalized linear models.
\newblock {\em Journal of the Royal Statistical Society. Series B\/}~{\em
  73\/}(1), 3--36.

\bibitem[\protect\citeauthoryear{Wood}{Wood}{2013}]{wood2013sp}
Wood, S.~N. (2013).
\newblock On p-values for smooth components of an extended generalized additive
  model.
\newblock {\em Biometrika\/}~{\em 100\/}(1), 221--228.

\bibitem[\protect\citeauthoryear{Wood}{Wood}{2017}]{wood2017igam}
Wood, S.~N. (2017).
\newblock {\em Generalized Additive Models: An Introduction with R\/} (2 ed.).
\newblock Chapman \& Hall/CRC.

\bibitem[\protect\citeauthoryear{Wood}{Wood}{2025}]{wood2025sup}
Wood, S.~N. (2025).
\newblock Supplement to ``On {N}eighbourhood {C}ross {V}alidation.''

\bibitem[\protect\citeauthoryear{Wood, Li, Shaddick, and Augustin}{Wood
  et~al.}{2017}]{wood2016giga}
Wood, S.~N., Z.~Li, G.~Shaddick, and N.~H. Augustin (2017).
\newblock Generalized additive models for gigadata: modelling the {UK} black
  smoke network daily data.
\newblock {\em Journal of the American Statistical Association\/}~{\em
  112\/}(519), 1199--1210.

\bibitem[\protect\citeauthoryear{Wood, Pya, and S{\"a}fken}{Wood
  et~al.}{2016}]{wood2015plig}
Wood, S.~N., N.~Pya, and B.~S{\"a}fken (2016).
\newblock Smoothing parameter and model selection for general smooth models
  (with discussion).
\newblock {\em Journal of the American Statistical Association\/}~{\em 111},
  1548--1575.

\bibitem[\protect\citeauthoryear{Xianyi, Qian, and Chothia}{Xianyi
  et~al.}{2014}]{openblas}
Xianyi, Z., W.~Qian, and Z.~Chothia (2014).
\newblock Open{BLAS}.
\newblock {\em URL: http://xianyi. github. io/OpenBLAS\/}.

\bibitem[\protect\citeauthoryear{Yee}{Yee}{2015}]{yee2015book}
Yee, T.~W. (2015).
\newblock {\em Vector {G}eneralized {L}inear and {A}dditive {M}odels: with an
  {I}mplementation in {R}}.
\newblock Springer.

\bibitem[\protect\citeauthoryear{Yee and Wild}{Yee and Wild}{1996}]{yee1996}
Yee, T.~W. and C.~Wild (1996).
\newblock Vector generalized additive models.
\newblock {\em Journal of the Royal Statistical Society. Series B\/}~{\em
  58\/}(3), 481--493.

\end{thebibliography}
%\bibliographystyle{chicago}

\else

\fi

\end{document}